\documentclass[fleqn,usenatbib]{mnras}

\usepackage{newtxtext,newtxmath}

\usepackage[T1]{fontenc}

\DeclareRobustCommand{\VAN}[3]{#2}
\let\VANthebibliography\thebibliography
\def\thebibliography{\DeclareRobustCommand{\VAN}[3]{##3}\VANthebibliography}

\usepackage{graphicx}	
\usepackage{amsmath}	

\usepackage{microtype}
\usepackage{rotating}
\usepackage{booktabs}
\usepackage{threeparttable}
\usepackage{tabularx}
\usepackage{subfigure}
\usepackage{xspace}

\newcommand{\project}[1]{\textsl{#1}\xspace}

\newcommand{\nicer}{\project{NICER}}

\newcommand{\nustar}{\project{NuSTAR}}
\newcommand{\extp}{\project{eXTP}}
\newcommand{\ixpe}{\project{IXPE}}
\newcommand{\grs}{GRS 1915+105\xspace}
\newcommand{\rms}{r.m.s.\xspace}

\newcommand{\deadtime}{\ensuremath{\tau_\mathrm{dead}}}

\title[Accurate Timing with SBI]{Accurate X-ray Timing in the Presence of Systematic Biases With Simulation-Based Inference}

\author[D.~Huppenkothen \& M.~Bachetti]{
Daniela Huppenkothen,$^{1}$\thanks{E-mail: d.huppenkothen@sron.nl}
Matteo Bachetti,$^{2}$
\\
$^{1}$SRON Netherlands Institute for Space Research, Niels Bohrweg 4, 2333 CA Leiden, The Netherlands \\
$^{2}$INAF-Osservatorio Astronomico di Cagliari, via della Scienza 5, I-09047 Selargius (CA), Italy
}

\date{Accepted XXX. Received YYY; in original form ZZZ}

\pubyear{2021}

\begin{document}
\label{firstpage}
\pagerange{\pageref{firstpage}--\pageref{lastpage}}
\maketitle

\begin{abstract}
Because many of our X-ray telescopes are optimized towards observing faint sources, observations of bright sources like X-ray binaries in outburst are often affected by instrumental biases. These effects include dead time and photon pile-up, which can dramatically change the statistical inference of physical parameters from these observations. While dead time is difficult to take into account  in a statistically consistent manner, simulating dead time-affected data is often straightforward. This structure makes the issue of inferring physical properties from dead time-affected observations fall into a class of problems common across many scientific disciplines. There is a growing number of methods to address them under the name of Simulation-Based Inference (SBI), aided by new developments in density estimation and statistical machine learning. In this paper, we introduce SBI as a principled way to infer variability properties from dead time-affected light curves. We use Sequential Neural Posterior Estimation to estimate the posterior probability for variability properties. We show that this method can recover variability parameters on simulated data even when dead time is variable, and present results of an application of this approach to NuSTAR observations of the galactic black hole X-ray binary GRS 1915+105.
\end{abstract}




\section{Introduction} \label{sec:intro}

As X-ray telescopes become more sensitive and detect even faint sources at a high signal-to-noise ratio, systematic biases become increasingly relevant over statistical fluctuations. Because many instruments are also optimized for faint sources, bright sources especially are targets where instrumental effects such as dead time and photon pile-up can significantly bias astrophysical inferences. 

Dead time is an instrumental effect whereby after arrival of a photon, a photon counting detector cannot record the arrival of a subsequent photon within a certain time interval. Dead time can be paralyzable or non-paralyzable. For paralyzable dead time, subsequent photons, while not measured, impose their own dead time on the detector, effectively prolonging the interval the detector cannot record events. In non-paralyzable dead time, photons arriving during this interval will be lost, but no longer affect the detector. 

For bright sources where a significant fraction of photons are lost, a paralyzable detector might be unable to record for a significant fraction of the time. Similarly, at high fluxes, a non-paralyzable detector may effectively record photons at a relatively regular interval set by the dead time. In both cases, the effect imposes a regular structure onto the resulting data set, which becomes readily apparent in Fourier representations of the light curve, and thus affects measurements of variability properties \citep{Zhang+95}. 

The effect of dead time on the periodogram can be easily modelled under two key assumptions: that (1) the variability of the observed astrophysical source is low, and (2) that the dead time for a given detector is constant \citep{vikhlininQuasiperiodicOscillationsShot1994,Zhang+95}. The latter implies that the dead time must be independent of photon energy and event grade, i.e.~the number and morphology of pixels involved in the detection of a single event.
However, usually neither of the two above conditions is fulfilled in practice.

Many of the sources we study using X-ray timing exhibit very strong stochastic or even flare-like variability \citep[e.g.][]{fender2004,walton2017}.
Moreover, the time a detector stops to process a given event is often driven by the number of pixels affected by the event, which is in turn related to the event grade and energy.
This can be a problem in missions where the dead time is comparable to the time scale of the physically interesting variability, like the Nuclear Spectroscopic Telescope Array \nustar (\citealt{harrison2013}, see \citealt{Bachetti+15}), and will be a problem in future missions like the Imaging X-ray Polarimetry Explorer (\ixpe, \citealt{ixpearxiv}; non-paralyzable dead time, 1.2 ms, priv. comm.), the enhanced X-ray Timing and Polarimetry mission's Large Area Detector (\extp/LAD, \citealt{extpspie}; non-paralyzable dead time, 83 $\mu$s, priv. comm.), and Athena's X-ray Integral Field Unit (X-IFU, \citealt{xifuspie}; paralyzable dead time, 1.2 ms, see \citealt{peillethesis})

\citet{Bachetti+15} takes advantage of the independence of \textit{NuSTAR}'s two identical detectors and proposes to use the \textit{cospectrum} as an alternative representation to traditional periodogram\footnote{The X-ray astronomy literature often uses the expression ``power density spectrum'' (PDS) or ``power spectral density'' (PSD) to refer to what is effectively a single realization of the PDS itself. Throughout this paper we will use the more proper term ``periodogram'' instead.} representations. Because dead time is a within-detector effect, its effects, unlike the intrinsic source variability in the light curve, are uncorrelated for identical detectors. The cospectrum effectively removes the amplitude variations imposed by dead time, though an imprint of the effect remains in that the noise variance of the cospectrum depends on frequency. 

In \citet{huppenkothen2018}, we formally derived the statistical distribution of the cospectrum for the limiting case of white noise, and found that the cospectrum is well-represented by a Laplace distribution. While this result is useful particularly for hypothesis testing against white noise (e.g.~to search for a strictly periodic signal from a pulsar or similar source), the presence of variability across a wide range of frequencies, as is commonly seen in accreting sources, severely distorts the statistical distribution of cospectral powers. As a result, the distributions derived in \citet{huppenkothen2018} are inappropriate for use as a likelihood when modelling broadband stochastic processes.

\citet{bachetti2018} effectively aims to circumvent that problem entirely, by yet again making use of independent, identical photon counting detectors present in telescopes like \textit{NuSTAR} and \textit{Fermi}'s Gamma-Ray Burst Monitor (GBM, \citealt{meegan2009}) in their \textit{Fourier Amplitude Differences} (FAD) method. It uses the differences between Fourier amplitudes to effectively build an empirical model for the way dead time affects the periodogram, and makes it possible to divide out most of the effect in both periodograms and cospectra. If the duration of the light curve is long compared to the timescales of the variability of interest, such that it is possible to generate Fourier representations averaged of $\sim 30$ or more individual segments, the problem can also be circumvented. In this limit, Gaussian distributions provide a reasonable approximation to periodograms, cospectra and related Fourier-based data representations like cross spectra, phase lags and the coherence \citep{huppenkothen2018,ingram2019}.

In this paper, we introduce a new formalism based on \textit{simulation-based inference} \citep[SBI; for a recent review, see][]{cranmer2020} that allows us to model the periodograms from instruments with complicated dead time distributions but lacking multiple detectors. The formalism is introduced---after a brief description of the data used in this paper (Section \ref{sec:dataprocessing})---in Section \ref{sec:sbi}.
This approach takes advantage of the specific structure of the problem we aim to solve: while it is difficult to build an accurate, general \textit{model} for dead time, it is relatively straightforward to \textit{simulate} from the process in many instances. This allows us to take advantage of recent advances in SBI to perform inference tasks as if we knew how to parametrize the dead time process as part of our model. 

In Section \ref{sec:simulations}, we extensively test this formalism on simulated light curves designed to mimic real-world observations taken with \textit{NuSTAR}, and also illustrate the generality of the approach laid out here to other problems with a similar structure, where instrumental effects impose biases onto the data that are difficult to model, but relatively straightforward to simulate. As an illustration, we show how the method can be used to accurately recover timing properties of simulated data designed to mimic an Active Galactic Nucleus (AGN) in the presence of gaps in the observations. (Section \ref{sec:agnbreaks}).
In Section \ref{sec:grs1915}, we demonstrate the method's validity on \textit{NuSTAR} observations of the galactic black hole X-ray binary GRS 1915+105. Finally, in Section \ref{sec:discussion}, we discuss the results, the limitations of the approach, and point out future applications.

\section{Data Processing}
\label{sec:dataprocessing}
We selected a public \nustar observation of the galactic black hole X-ray binary \grs with known quasi-periodic oscillations (QPOs) \citep{ShreeramIngram2020}. 
The observation, ObsID 80401312002, was part of a target-of-opportunity joint \nicer-\nustar program to study the QPOs from the source, and was executed on UT 2018-06-08.
We downloaded the data from the High Energy Astrophysics Science Archive Research Center (HEASARC) and processed them with the \texttt{nupipeline} FTOOL shipped with HEAsoft 6.28, using the default options.
We barycentered the data using the known optical position of the source \citep{2003yCat.2246....0C}, the International Celestial Reference System (ICRS) reference frame and the Planetary and Lunar Ephemeris DE 421.

We extracted events from a region of 70" centered on the position of the source on the detector. 
Due to small astrometric differences between the two detectors, we adjusted and centered the source position with the \texttt{peak\_local\_max} function in Scikit-Image v. 0.18 \citep{scikit-image} using images in detector coordinates.

\section{Simulation-Based Inference}
\label{sec:sbi}

The aim of statistical inference, particularly in a Bayesian context, is to infer causal knowledge about physical processes from observed data. This process requires several components. First, it requires a \textit{generative model}: a function $f(x,\theta)$ that will generate values $y_\mathrm{model,i} = f(x,\theta)$ that are assumed to be a reasonable approximation of the real process presumed to have generated the data. This function is specified by some parameters $\theta$ governing the shape of $f$, given some dependent variable $x$ (e.g. time of observations, wavelength, spatial coordinates). The data vector $y_\mathrm{obs}$ and model are compared via a \textit{likelihood function} $\mathcal{L}(\theta) = p(y_\mathrm{obs} | \theta)$, an analytical relationship determined by the measurement process: for many astronomical instruments measuring large incident fluxes (e.g.~in optical astronomy), a Gaussian likelihood is often assumed to be valid. In X-ray astronomy, where individual incident photons are recorded, a Poisson distribution is generally appropriate for light curves, whereas a $\chi^2$ likelihood is applicable to Fourier periodograms. Inference of the parameters $\theta$ then proceeds via Bayes' rule:

\begin{equation}
p(\theta | y_\mathrm{obs}, I) \propto p(y_\mathrm{obs} | \theta, I) p(\theta | I) \; ,
\end{equation}

\noindent where$p(\theta | I)$ describes the prior knowledge and constraints on parameters $\theta$ before inference, and $I$ encapsulates all inherent assumptions decisions made in setting up the model (e.g. the shape of the prior probability distribution, the form of the likelihood). The posterior probability density $p(\theta | y_\mathrm{obs})$ is generally not analytically tractable except in simple problems, and thus is often numerically approximated through methods like Markov Chain Monte Carlo \citep{metropolis1953,hastings1970} or Nested Sampling \citep{skilling2004}. 

There are two inherent assumptions in this process that may not be true in practice: (1) the function $f(x,\theta)$ giving rise to the emission can be directly compared to the observed data points, and (2) it is possible to write down an analytical form of the likelihood. The first assumption might be challenged when the process in question is inherently stochastic. Because the observed data is a single realization of a process with sometimes infinite possible realizations, comparing data to the model directly is not possible. In X-ray spectral timing, this problem is often circumvented by modelling summarizing representations of the data that approximate the process from which the realization was drawn (e.g.~Fourier periodograms, the \rms-flux relation describing the relationship between root-mean-square variability and brightness) rather than the raw data. The second assumption might be broken for example when the measurement process is complicated and as a consequence the likelihood becomes intractable. Spectral timing of bright accreting sources is complicated by both: the underlying process generating the X-ray flux we observe is stochastic, and for many instruments, dead time heavily censors the observed list of photon arrivals. Because dead time directly depends on the incident flux, an analytical model is only available in the simplifying case where dead time is constant and does not depend on energy or event grade. The \textit{cospectrum} can be a powerful alternative for problems that involve tests against white noise (e.g. searching for pulsars), but has an analytically intractable likelihood when stochastic processes contribute to the variable incident flux (\citealt{huppenkothen2018}, Huppenkothen \& Bachetti in prep.). For instruments without multiple detectors, the cospectrum may simply not be available. 

The intractability of the likelihood for cospectra removes one important avenue for parametric modelling of stochastic variability in Fourier space in the presence of dead time. Without a likelihood, parametric models of QPOs and other stochastic processes cannot be compared to the cospectrum. Even with a likelihood, the effects of dead time would not be completely removed from the cospectrum: the imprints of dead time are still observable in the \textit{variance} of the noise distributions, and might hence bias the resulting parameter inference for physical components inferred to be present in the light curve.

Modelling the periodogram in the presence of dead time instead presents its own challenges: while the periodogram retains its fundamental $\chi^2$ properties, dead time imprints its effects strongly onto the underlying power spectrum. Parametric models for dead time exist, but dead time scales strongly with the flux impinging on the instrument: intervals of high source flux will be more strongly affected by dead time than intervals of low flux. Because the objects we study tend to vary rapidly, sometimes over multiple orders of magnitude, and variability tends to be driven by stochastic processes, quantifying the effect of dead time on a specific observation is difficult: in statistical terms, a precise estimate of the dead time would involve an integral over \textit{all possible} realizations of the underlying stochastic process:

\[
p(y_\mathrm{obs} | \theta) = \int{dz\, p(y_\mathrm{obs}, z | \theta)}
\]

\noindent where $y_\mathrm{obs}$ is the observed periodogram, $\theta$ are input parameters to the underlying variability model, and $z$ are all possible realizations of the stochastic process that could have produced $y_\mathrm{obs}$. 

However, for many detectors, dead time is not difficult to simulate. If the dead time of the detector is well-known, one can draw individual photons from a Poisson process given some underlying varying mean Poisson rate $\lambda$, and remove those photons that arrive within the dead time interval of a previous photon (in the non-paralyzable case, though equivalent simulation routines exist for paralyzable dead time; e.g.~\citealt{Zhang+95}. Both are implemented in the software package \textit{stingray}; \citealt{stingraypaper}). While in a traditional likelihood, one compares an underlying, noiseless \textit{function} to an observed, noisy \textit{observation}, one now faces the problem of comparing two noisy data sets: one observed, one simulated. Because both the underlying process that generated the data and the measurement process itself are stochastic, a direct comparison of the observed and simulated data sets is difficult. 

The structure of this problem, where a likelihood does not exist or is implicitly defined in a simulator, but simulations of the full process are relatively easily accessible, is sometimes called \textit{likelihood-free inference} (LFI) and is common in science. Early solutions were first suggested by \citet{diggle1984monte} and \citet{rubin1984bayesianly} and later formulated in the context of population ecology under the name \textit{Approximate Bayesian Computation} (ABC; \citealt{beaumont2002approximate}. These early works show that it is possible to accurately recover a Bayesian posterior probability distribution from simulations, given a well-designed metric to measure the distance between the observed and simulated data. The traditional ABC rejection sampling algorithm is remarkably straightforward \citep{tavare1997inferring,pritchard1999population}. First, draw parameter sets from the prior. Then, for each parameter set, generate an observation incorporating all measurement effects and biases that the real observed data is believed to be affected by. Generally, simulated and observed data sets may be too noisy and high-dimensional for direct comparison, so they are often transformed into summary statistics, designed to encode as much information relevant to the inference process as possible. Each simulation is then compared to the observed data through a \textit{distance metric} (e.g.~the Euclidean distance, though many other metrics exist): all parameter sets for which the distance between simulations and the observed data is below a specified threshold $\epsilon$ are kept, all others are discarded. This approach suffers from a natural trade-off between the desire to be accurate (i.e.~making $\epsilon$ very small), and the computational feasibility of generating orders of magnitude more observations that are rejected than those that are kept. 

Subsequent work improved on all aspects of this approach in order to make it computationally feasible, including more efficient sampling methods, better distance metrics and alternative, non-Bayesian formulations. Recent work, often under the name \textit{simulation-based inference} (SBI\footnote{In many contexts the respective terms ABC, LFI and SBI can be and are used interchangeably. In the rest of this paper, we will follow recent convention and use the term SBI.}), combines the ideas of ABC with advances in machine learning in order to circumvent some of the traditional ABC problems through the use of neural networks. The latter can learn to emulate the simulation process, but can also be used to directly learn the likelihood \citep{papamakarios2019sequential,lueckmann2019likelihood}, likelihood ratios 
\citep{hermans2019likelihood,durkan2020contrastive} or posterior density \citep{papamakarios2016fast,2017arXiv171101861L,greenberg2019}. For recent reviews, see \citet{sisson2018handbook} and \citet{cranmer2020}, and a benchmarking of recent algorithms can be found in \citet{2021arXiv210104653L}. These methods take advantage of recent advances in using neural networks to specify flexible probability distributions. 

Neural Posterior Estimation (NPE; \citealt{npe1,npe2,2017arXiv171101861L,npe3,npe4}) uses this approach to directly infer the target posterior. In its original version, NPE uses a large set of simulated data sets drawn from the prior to learn a mapping between simulator output $\hat{y}$ and (potentially complex, multi-modal) posteriors $p(\theta | \hat{y})$. The posterior estimate is selected from a family of densities $q_\psi$, where $\psi$ are distribution parameters and generally the output of a neural network $F(\hat{y}, \phi)$ with neural network weights $\phi$. These weights are learned by minimizing a form of the negative log-likelihood, $\mathcal{L} = - \sum_{j=1}^{N}{\log{q_{F(\hat{y}_j, \phi)}(\theta)}}$. The trained proposal posterior probability density approximates the true posterior, $q_{F(\hat{y}_j, \phi)}(\theta) \approx p(\theta | \hat{y})$. This might require a reasonably large set of simulations in order to fully map out the posterior space. However, it has the advantage of \textit{amortization}: once the network is trained, drawing posterior samples is very fast, and can be done for multiple independent observations using the same model without retraining. The sequential version of this algorithm (Sequential Neural Posterior Estimation; SNPE) is based on the insight that for inference on a given data set, the prior is an inefficient proposal distribution for generating simulated data. In this version, a network is trained on a small set of simulations, and the posterior constructed for the observed data. This posterior is then used as the proposal distribution $\Tilde{p}(\theta)$ for simulations in the next round. \citet{papamakarios2016fast} show that this drastically reduces the number of simulations required, but loses its amortization. Because in SNPE the proposal distribution for subsequent inference rounds is conditioned on a single set of observed data, the trained network can no longer be used to infer the posterior for other data sets. In addition, it requires either post-processing of the distribution (or its samples), or a reformulation of the loss function (i.e.~the function describing the distance between the approximate distribution and the true posterior), because minimizing $\mathcal{L}$ no longer yields the target posterior, but rather a \textit{proposal posterior} $\Tilde{p}(\theta | \hat{y})$. \citet{greenberg2019} suggest two improvements: (1) a new parametrization of the loss function that enables automatic transformation between estimates of $p(\theta | \hat{y})$ and $\Tilde{p}(\theta | \hat{y})$, and (2) ``atomic'' proposals. The former change enables the recovery of the ground truth posterior $p(\theta | \hat{y})$ during inference. The latter allows an arbitrary choice of density estimators, priors and posteriors.

Problems of this type, where a simulator is accessible and computationally feasible, but a likelihood is not, are common in astronomy, and have led to a recent rise in work implementing SBI models, in a wide range of areas including cosmology \citep{jennings2016,jennings2017,2017JCAP...08..035H,2017MNRAS.469.2791H,2020MNRAS.493.5913L,2019MNRAS.490.4237L,2019MNRAS.488.4440A,2020PhRvD.101h2003K}, solar physics \citep{2021ApJS..252....9W}, supermassive black holes \citep{2020arXiv201109582W}, exoplanets \citep{2019MNRAS.489.3162S,2020MNRAS.498.2249H,2020AJ....160..200B,2021AJ....161...69K}, stellar astronomy \citep{2019arXiv190411306C,2020AJ....159..248K,2020ApJ...893...67M}, galactic astronomy \citep{2019A&A...624L...1M,2020arXiv201009721C}, dark matter studies \citep{2020arXiv201114923H}, and extragalactic astronomy \citep{2020A&A...635A.136A,2020JCAP...09..048T,2020arXiv201013221H,2020MNRAS.496.1718E}. 

Our goal here is to implement an SBI model with SNPE for dead time in X-ray detectors. We first build a simulator for \nustar-like dead time, and generate periodograms with a quasi-periodic oscillation in order to study how well SNPE can recover the parameters used to generate the data. \nustar~presents a convenient test case, because alternative methods taking advantage of its two independent detectors exist and can be used to compare performance. We explore both amortized and sequential versions of the neural posterior estimation algorithm with automatic posterior transformation as introduced by \citet{greenberg2019} and implemented in the Python package \texttt{sbi} \citep{Tejero-Cantero2020}. As a density estimator we choose a Masked Autoregressive Flow (MAF; \citealt{papamakarios2017}). MAFs combine two recent ideas: autoregressive density estimation and normalizing flows. Autoregressive density estimation decomposes the joint density $p(\theta | \hat{y})$ into the product of one-dimensional conditional densities: $p(\theta | \hat{y}) = \prod_{k}{p(\theta_k | \theta_{1:k-1}, \hat{y})}$, where these conditional densities are generally simple one-dimensional distributions (e.g. a Gaussian) with parameters estimated through a neural network conditioned on all previous $\theta_{1:k-1}$. This ascribes intrinsic meaning to the ordering of the parameters $\theta$, which may or may not be realistic in practical situations. Normalizing flows, on the other hand, are based on the idea of taking a base distribution, and applying a sequence of $m$ invertible transforms (with learnable parameters) to generate a more complex target distribution. \citet{papamakarios2017} suggested to stack multiple subsequent autoregressive transformations, with each implementing a different ordering for $\theta$ in order to construct a MAF. Here, we use a MAF with 5 transformation and 50 hidden units each, and embed the MAF in both the sequential and amortized versions of the NPE algorithm as explained above.

\section{Simulations with Dead Time}
\label{sec:simulations}

We present results for SBI in the form of (S)NPE applied to simulated light curves and their resulting periodograms. We set up a realistic toy problem where the goal is accurate inference of the centroid frequency, width and fractional \rms amplitude of a quasi-periodic oscillation (QPO) in the presence of dead time. We simulate both a low-frequency QPO as well as a high-frequency QPO, and we explore both inference on single periodograms as well as averaged periodograms\footnote{Pre-run simulations to train the neural networks are available on \href{http://doi.org/10.5281/zenodo.4670161}{Zenodo}. All code related to the project, including notebooks detailing the simulation and inference procedure, as well as Python scripts to run simulations, can be found \href{https://github.com/dhuppenkothen/DeadTimeSBI}{in the GitHub repository}}. For averaged periodograms, we compare our results to posterior distributions sampled with a Markov Chain Monte Carlo approach with a $\chi^2$ likelihood. In the latter case, periodograms were corrected for dead time using the Fourier Amplitude Differencing (FAD) method. 

\begin{figure}
\begin{center}
\includegraphics[width=0.5\textwidth]{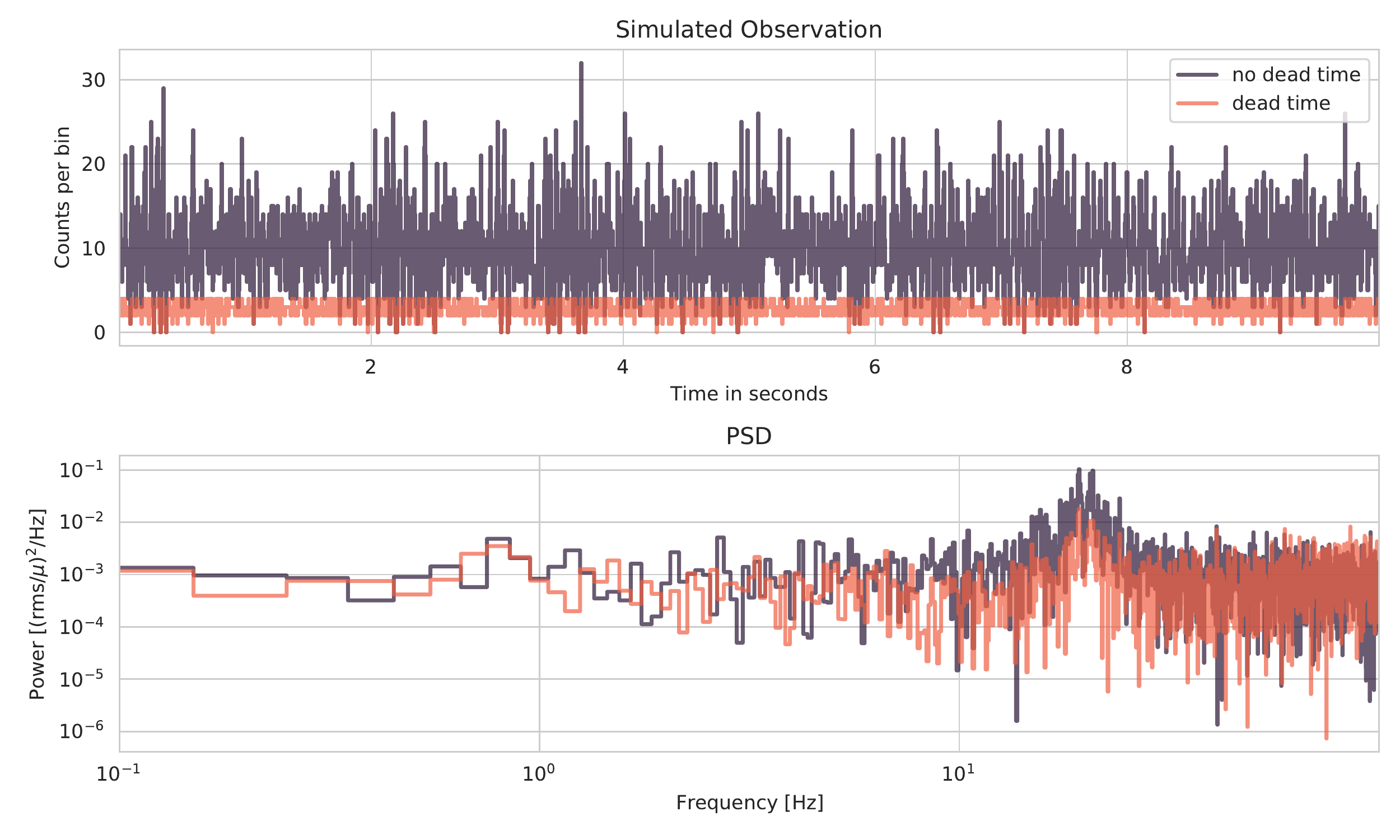}
\caption{Simulated light curve and Fourier products for a single QPO at $20\mathrm{Hz}$. Top panel: simulated light curves both with (orange) and without (purple) dead time applied. Bottom panel: periodograms corresponding to the light curves in the top panel.}
\label{fig:lf_data}
\end{center}
\end{figure}

\begin{table*}[hbtp]
\renewcommand{\arraystretch}{1.3}
\footnotesize
\caption{Priors used in the models}
\begin{threeparttable} 
\begin{tabularx}{18cm}{p{6.5cm}p{2.5cm}p{4.5cm}p{4.5cm}}
\toprule
 \bf{Model} & \bf{Parameter} & \bf{Meaning} & \bf{Probability Distribution} \\ \midrule
{\it SBI Models, shared parameters} & $\mathrm{rms}_f$ & fractional \rms amplitude & $\mathrm{Uniform}(0.1,0.5)$ \\
& $q$ & QPO quality factor & $\mathrm{Uniform}(3,30)$ \\
& $\mu_\mathrm{cr}$ & mean incident count rate & $\mathrm{Uniform}(500,1500)$ \\ \midrule
{\it SBI Models, LF QPO simulations} & $\nu_0$ & QPO centroid frequency [Hz] & $\mathrm{Uniform}(5, 40)$ \\ \midrule
{\it SBI Models, HF QPO simulations, single periodogram} & $\nu_0$ & QPO centroid frequency [Hz] & $\mathrm{Uniform}(100, 300)$ \\ \midrule
{\it SBI Models, HF QPO simulations, averaged periodogram} & $\nu_0$ & QPO centroid frequency [Hz] & $\mathrm{Uniform}(100, 500)$ \\ \midrule
{\it Traditional Bayesian Models, Shared parameters} & $A_\mathrm{QPO}$ & QPO amplitude & $\mathrm{Uniform}(10^{-10}, 100)$ \\
& $\Delta \nu$ & QPO FWHM & $\mathrm{Uniform}(0.01,40)$ \\
& $A_\mathrm{wn}$ & white noise amplitude &  $\mathrm{Uniform}(10^{-20},10^5)$ \\ \midrule
{\it Traditional Bayesian Models, LF simulations} & $\nu_0$ & QPO centroid frequency [Hz ]& $\mathrm{Uniform}(5, 50)$ \\ \midrule
{\it Traditional Bayesian Models, HF simulations} & $\nu_0$ & QPO centroid frequency [Hz] & $\mathrm{Uniform}(100, 500)$ \\

\bottomrule
\end{tabularx}
   \begin{tablenotes}
      \item{An overview over the model parameters with their respective prior probability distributions for the models in this section.}
\end{tablenotes}
\end{threeparttable}
\label{tab:priors}
\end{table*}

\subsection{Single Periodogram: Low-Frequency (LF) QPO}
\label{sec:single_lf}

For simulating light curves with a single QPO, we assume an underlying power spectrum consisting of a single Lorentzian with a centroid frequency of $\nu_0 = 20\mathrm{Hz}$, a quality factor $q=10$ (resulting in a narrow peak with a full width at half maximum of $\Delta\nu = 2\mathrm{Hz}$). 
We assume a fractional \rms amplitude of 0.4 for the signal. Given these parameters, we simulate a single light curve of length $T = 10\mathrm{s}$ with a high time resolution of $dt = 10^{-5}\mathrm{s}$ based on the underlying power spectrum using the method of \citet{timmer1995}. 
The high time resolution is chosen to be significantly better than the average dead time in \nustar, $dt_\mathrm{dead}=0.0025\mathrm{s}$ in order to allow for an effective simulation of the dead time process. 
We rescale the simulated light curve such that the mean incident count rate is $1000\, \mathrm{counts}/\mathrm{s}$.
The count rate is chosen to be high enough for dead time to have an appreciable impact. In order to turn the simulated light curve into events, we first draw from a Poisson distribution for each bin in the light curve, using the simulated rate in that bin as a Poisson parameter. We then take all time bins that contain more than one count and randomly distribute $n$ events within the time bin according to a uniform distribution, where $n$ is the number of counts in that bin. 
We repeat this process to generate two light curves, which are identical except for the Poisson noise. This effectively simulates the behaviour of observing a single source with two independent \nustar~detector modules, \textit{Focal Plane Module A} (FPMA) and \textit{Focal Plane Module B} (FPMB).
To the resulting event lists we apply \nustar-like, non-paralyzable dead time: all photons that arrive within $0.0025\,\mathrm{s}$ of the previous photon are discarded as having arrived while the detector was unresponsive. The resulting event lists are treated as an observation. We generate light curves at a coarser time resolution of $dt_\mathrm{obs} = 0.005$\,s, since we are primarily interested in a QPO at $20\mathrm{Hz}$ and can ignore the higher frequencies, and sum the two light curves together. 
We present the simulated light curve--both with and without dead time--along with the associated periodograms in Figure \ref{fig:lf_data}. Because this simulated light curve corresponds to an observation of an extraordinarily bright source, dead time has a very strong effect on the measured counts. The average count rate in the dead time-affected light curve is $550\,\mathrm{counts}/\mathrm{s}$ compared to $2000\,\mathrm{counts}/\mathrm{s}$ (for a sum of two light curves from independent detectors and an incident rate of $1000\,\mathrm{counts}/\mathrm{s}$). This corresponds to a loss of about 72\% of incident photons. 

\begin{figure}
\begin{center}
\includegraphics[width=0.5\textwidth]{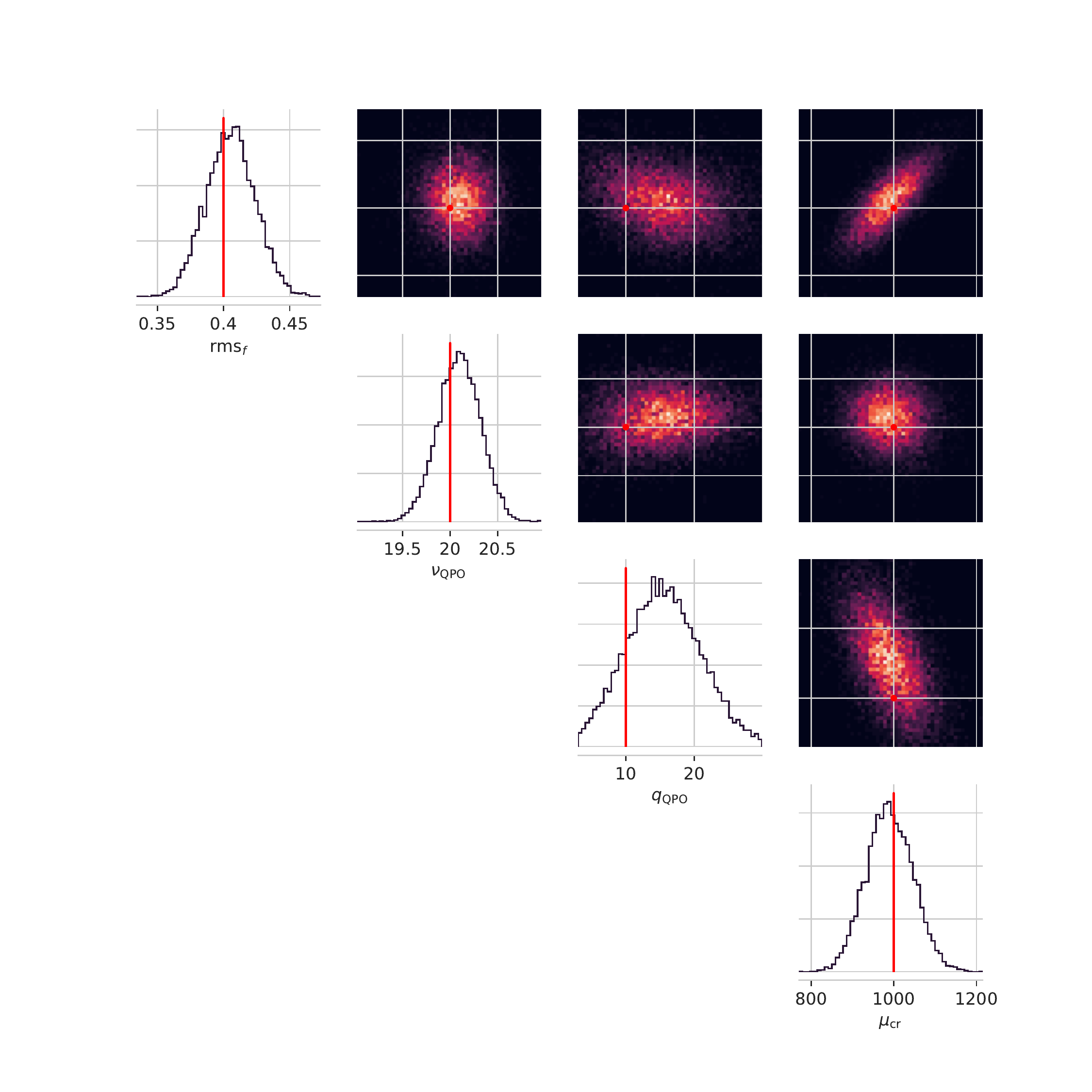}
\caption{Posterior distribution as derived through amortized SNPE: on the diagonal, we show one-dimensional marginalized posterior densities, on the off-diagonal a heat map of parameter pairs. All distributions are normalized so that they integrate to one. In red, we mark the true parameters that generated the data. For all parameters, the posterior clusters tightly around the true value.}
\label{fig:lf_corner}
\end{center}
\end{figure}

\begin{figure}
\begin{center}
\includegraphics[width=0.5\textwidth]{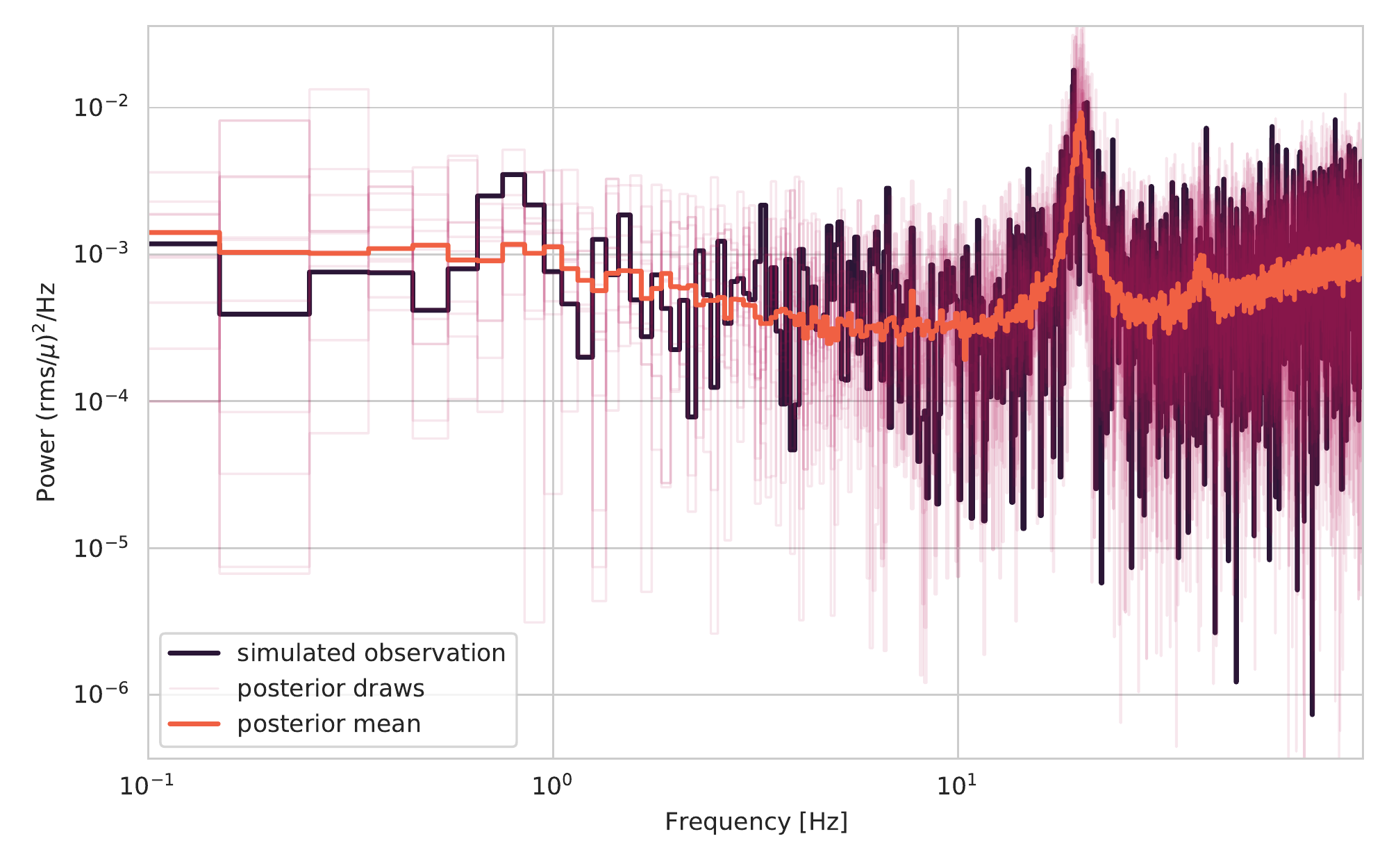}
\caption{We show the simulated observation (dark purple), along with 100 random draws from the posterior (light pink), as well as the posterior median derived from these 100 random draws (orange). The draws from the posterior clearly trace out the QPO. In addition, the posterior median makes frequency-dependent changes in the white noise level due to dead time evident.}
\label{fig:lf_post}
\end{center}
\end{figure}

\begin{figure*}
\begin{center}
\includegraphics[width=\textwidth]{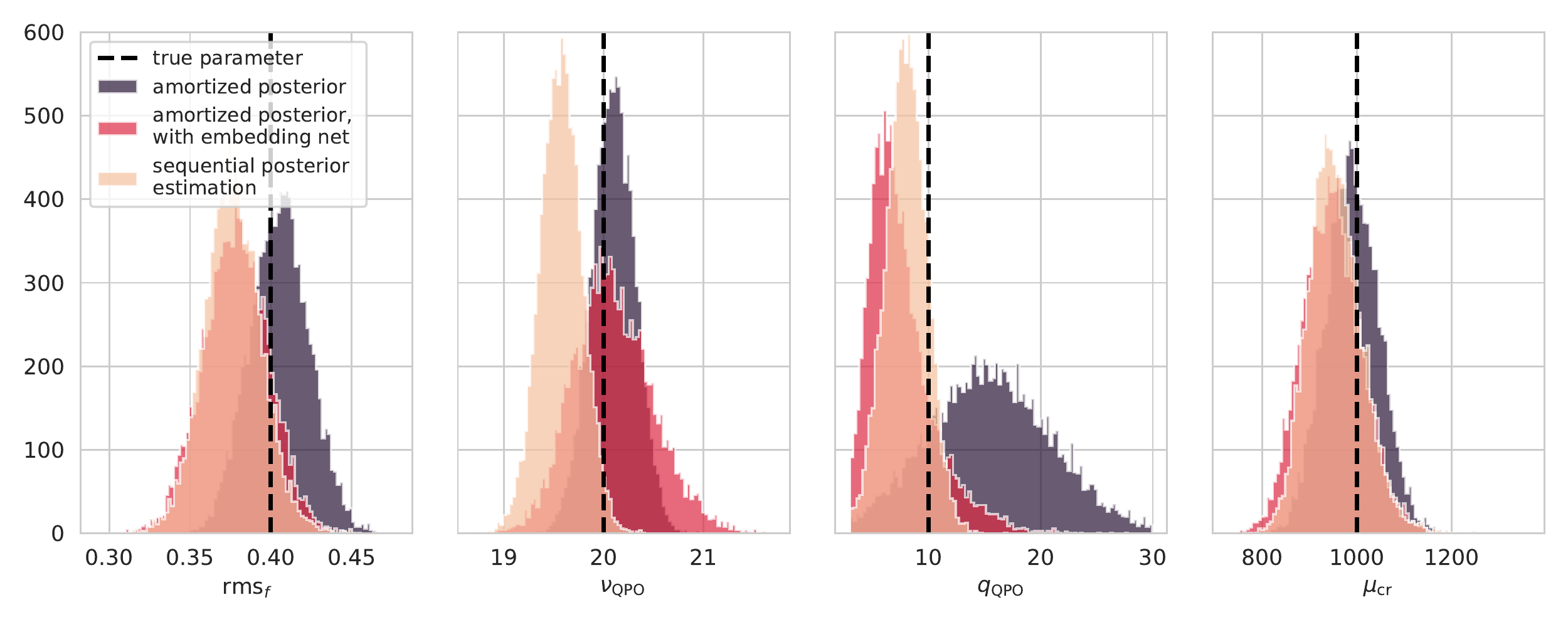}
\caption{Posterior probability distributions for three different SBI models: (1) amortized posterior inference with 50,000 simulations on the raw periodogram (purple), (2) amortized posterior inference with 50,000 simulations, with an embedding net generating summaries of the data (pink), (3) sequential posterior estimation on the raw periodogram, for five rounds of 1,000 simulations each (orange). }
\label{fig:lf_params_comp}
\end{center}
\end{figure*}

The goal of the SBI procedure is to accurately infer the parameters of the QPO present in the simulated data. We aim to infer four parameters: the QPO centroid frequency, $\nu_0$; the quality factor of the QPO $q$, related to the full width at half-maximum, $\Delta \nu$, through the equation $q = \nu_0 / \Delta \nu$; the fractional \rms amplitude of the variability present in the QPO, $\mathrm{rms}_f$; and the incident mean count rate without dead time, $\mu_\mathrm{cr}$. For simplicity we set flat priors for all parameters in our toy experiments, with bounds defined in Table \ref{tab:priors}. Because our data is very informative (the fractional rms amplitude of the QPO is very high), we expect the prior choice to have little impact on the results for the toy problem set out here. Our initial experiments suggest that choosing very wide, uninformative priors for the QPO centroid frequency (allowing this frequency to range over multiple orders of magnitude) requires a higher number of simulations in order to cover enough of the available parameter space. How this increased computational requirement manifests in practice depends on the other choices made in setting up the problem (e.g.~number of model components, priors for other parameters).

We then generate $50,000$ parameter sets from the prior, and generate simulated periodograms assuming \nustar-like dead time, using the same procedure we used to generate our ``observation''. For the first inference round, we use all $50,000$ simulated periodograms to build a Masked Autoregressive Flow that approximates the posterior probability density, allowing for amortized inference. We use the raw periodogram as a feature vector. Because the periodogram is very noisy, and SBI procedures generally perform best on well-designed summary statistics of the data, we would naively not expect good performance when comparing raw data sets.

We sample from this posterior, and present the results in Figures \ref{fig:lf_corner} and \ref{fig:lf_post}. We find that even when dead time removes a large fraction of the incident photons and we use the raw periodogram as features, the SBI procedure is capable of recovering all four parameters with a high degree of accuracy: as the corner plot shows, the posterior probability density clusters relatively tightly around the true input parameters.  

Because the QPO is strong, accurate inferences for the centroid frequency and quality factor are not surprising. It is noteworthy, however, that the model accurately recovers both the fractional \rms amplitude as well as the incident mean count rate, because we limited the periodogram here to a Nyquist frequency of $100 \mathrm{Hz}$ for computational efficiency. This means that there is enough information about the dead time present at low frequencies to accurately infer the true incident count rate, despite the model not having access to the full imprint of dead time onto the periodogram at higher frequencies. This is possibly caused by the dead time-dependent drop in the white noise level at low frequencies. 

In a second experiment, we repeat the inference process with the same $50,000$ simulations, but include a convolutional neural network to generate summaries of the data. We train multiple such \textit{embedding networks} with different architectures, though all contain at least  convolutional layer, a max-pooling operation and a fully connected layer. We use Rectified Linear Unit (ReLU) activation functions, and trained architectures with both one and two convolutional layers. The embedding net is trained concurrently with the density estimator, using the same loss function. In all cases, we recover at most comparable performance to using the raw periodogram. Most embedding networks produce biased results, and lose precision especially for the quality factor (Figure \ref{fig:lf_params_comp}). We settle on an architecture with 1 convolutional layer, a kernel size of 12 and a set of 12 output summary features for subsequent experiments.

In a final experiment, we use the sequential learning process of SNPE: instead of learning an approximation to the posterior density from a single set of $50000$ simulations, we run $r=5$ rounds of inference with $N=1000$ simulations each, and train a MAF in each round.
We find that using SNPE, we can generate a posterior probability density with comparable precision and accuracy as that derived from the full $50000$ simulations using only the five rounds of $1000$ simulations each, or $5000$ simulations total (Figure \ref{fig:lf_params_comp}). This amounts to a speed-up of a factor ten compared to the amortized inference process, which is especially helpful when the data set of interest consists of a single data set. As we will show in Section \ref{sec:grs1915} below, amortized inference, in comparison, can be very efficient when inferences are to be made over a large series of periodograms. We quantify the performance of the three models--the amortized version, the sequential version, and the model with an embedding net--by calculating the distance between the mean of the posterior distribution and the known true parameter value. In general, all three models perform acceptably at recovering the true parameters. In all but two cases, the true value is within 1 standard deviation of the mean of the posterior distribution. Exceptions are the posterior for $\nu_\mathrm{QPO}$ in the amortized model ($\Delta_{\nu_\mathrm{QPO}} = 1.48 \sigma$) and the posterior for $\mathrm{rms}_r$ in the model with the embedding net ($\Delta_{\mathrm{rms}_r} = 1.36 \sigma$). However, given the large uncertainties inherent in the periodogram estimator, and the additional loss of information due to the removal of photons because of dead time, these deviations are within acceptable tolerances. The sequential model outperforms the other two models on all parameters except the fractional rms amplitude. The distances for all three models are summarized in Table \ref{tab:lf_post_distance}. 

\begin{table}
\centering{
\renewcommand{\arraystretch}{1.3}
\footnotesize
\caption{LF QPO: Distance between true parameters and posterior mean}
\begin{threeparttable} 
\begin{tabularx}{8.0cm}{p{1.5cm}p{1.0cm}p{1.0cm}p{1.0cm}p{1.0cm}}
\toprule
\bf{Parameter} & \bf{True value} & $\Delta_\mathrm{am}$ & $\Delta_\mathrm{EN}$ & $\Delta_\mathrm{SE}$ \\ \midrule
 $\mathrm{rms}_f$   & $0.4$     & $0.05\sigma$  &  $1.36\sigma$ & $0.55\sigma$ \\
 $\Delta\nu_0$      & $20.0$    & $1.48\sigma$  &  $0.46\sigma$ & $0.20\sigma$ \\
 $q_\mathrm{QPO}$   & $10.0$    & $0.38\sigma$  &  $0.85\sigma$ & $0.17\sigma$ \\
 $\mu_\mathrm{cr}$  & $1000.0$  & $0.18\sigma$  &  $0.85\sigma$ & $0.16\sigma$ \\
\bottomrule
\end{tabularx}
   \begin{tablenotes}
      \item{We provide a rough quantification of how well the posterior distributions recover the true parameter values through calculating the distance between the true parameters and the posterior mean, in units of standard deviations from the mean $\sigma$, for the three models (AN: amortized model; EN: model with embedding net); SE: sequential model.}
\end{tablenotes}
\end{threeparttable}
\label{tab:lf_post_distance}}
\end{table}

We compare the SBI process to an ordinary model using the standard $\chi^2$ likelihood for periodograms without taking dead time into account. Because the QPO has a relatively low frequency where dead time effects might be less pronounced, one might expect that this model would generate comparable or better results, which in turn would have advantages in terms of computationally efficiency. We build a model for the periodogram using a single Lorentzian to represent the QPO, and, where possible, implement the same priors as for the SBI model. Note, however, than instead of parameters for the fractional \rms amplitude and the mean count rate, the periodogram model here includes an amplitude for the Lorentzian function and a white noise level in the periodogram (priors for these parameters are also included in Table \ref{tab:priors}). We then sample from the posterior using Markov Chain Monte Carlo as implemented in the Python package \textit{emcee} \citep{emcee}, and compare the posterior distribution computed with SNPE to the version sampled using MCMC. 

\begin{figure}
\begin{center}
\includegraphics[width=0.5\textwidth]{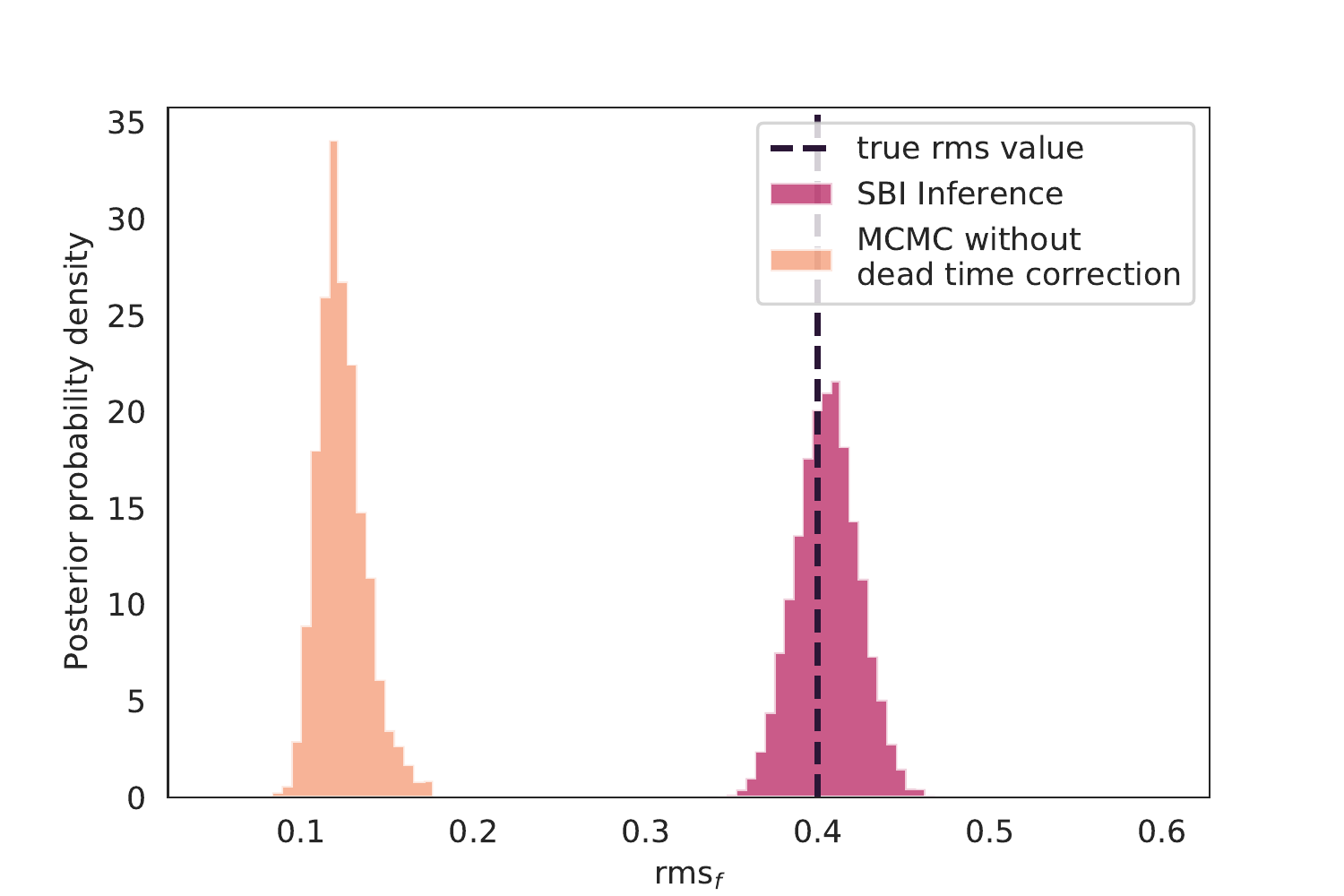}
\caption{We compare the fractional \rms amplitude derived from the SBI model (pink) with the traditional Bayesian model sampled with MCMC (orange). While the SBI model directly sampled the fractional \rms amplitude, we calculated the distribution of $\mathrm{rms}_f$ for the Bayesian model by integrating under the white-noise corrected, \rms-normalized model periodogram for 1000 parameter sets drawn from the posterior. While the Bayesian model can approximate the centroid frequency and the width of the QPO well, it fails estimating the fractional \rms amplitude correctly in the presence of dead time, because the latter removes events especially when the flux is high. This, in turn, leads to a lower variance in the light curve than without the effects of dead time.}
\label{fig:lf_rms_comp}
\end{center}
\end{figure}

In Figure \ref{fig:lf_rms_comp}, we compare posteriors for the fractional \rms amplitude derived with both approaches. We find that while the SNPE model accurately recovers the injected \rms amplitude, the likelihood-based model without dead time leads to a highly biased posterior probability density which underpredicts the fractional \rms amplitude by a factor of four, with a high degree of confidence (the true value is excluded at the $12
\sigma$ level). This can largely be explained by the flux-dependent loss of photons due to dead time. More photons are lost to dead time when the incident flux is high. This, in effect, reduces the intrinsic variability in the light curve, and leads to a lower inferred fractional \rms amplitude \citep{vanderklis1989,Bachetti+15}.

\begin{figure}
\begin{center}
\includegraphics[width=0.5\textwidth]{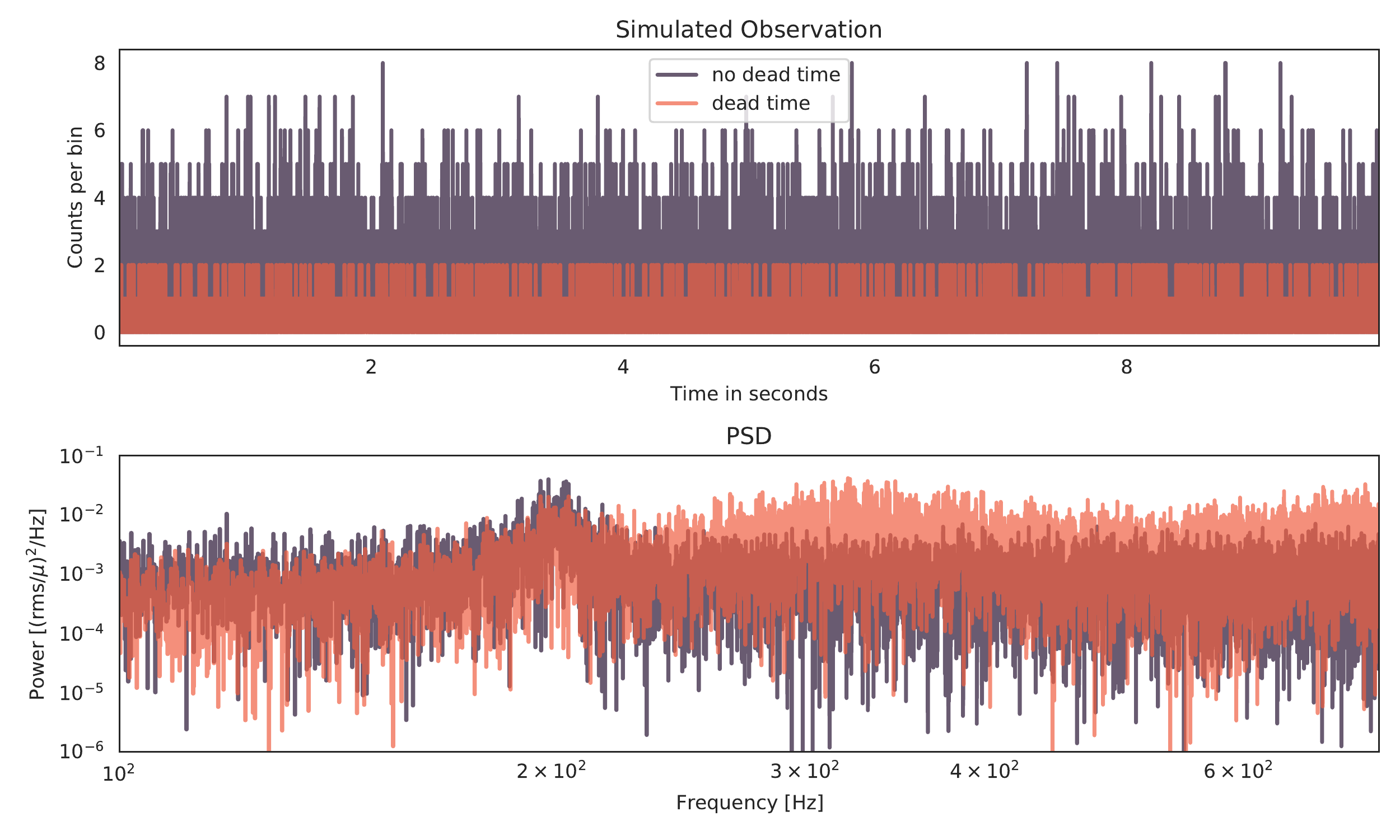}
\caption{Simulated light curve and Fourier products for a single QPO at $200\mathrm{Hz}$. Top panel: simulated light curves both with (orange) and without (purple) dead time applied. Bottom panel: periodograms corresponding to the light curves in the top panel. The effects of dead time in the periodogram are apparent at high frequencies, where there is additional power in a very broad, wavy pattern.}
\label{fig:hf_data}
\end{center}
\end{figure}

\begin{figure}
\begin{center}
\includegraphics[width=0.5\textwidth]{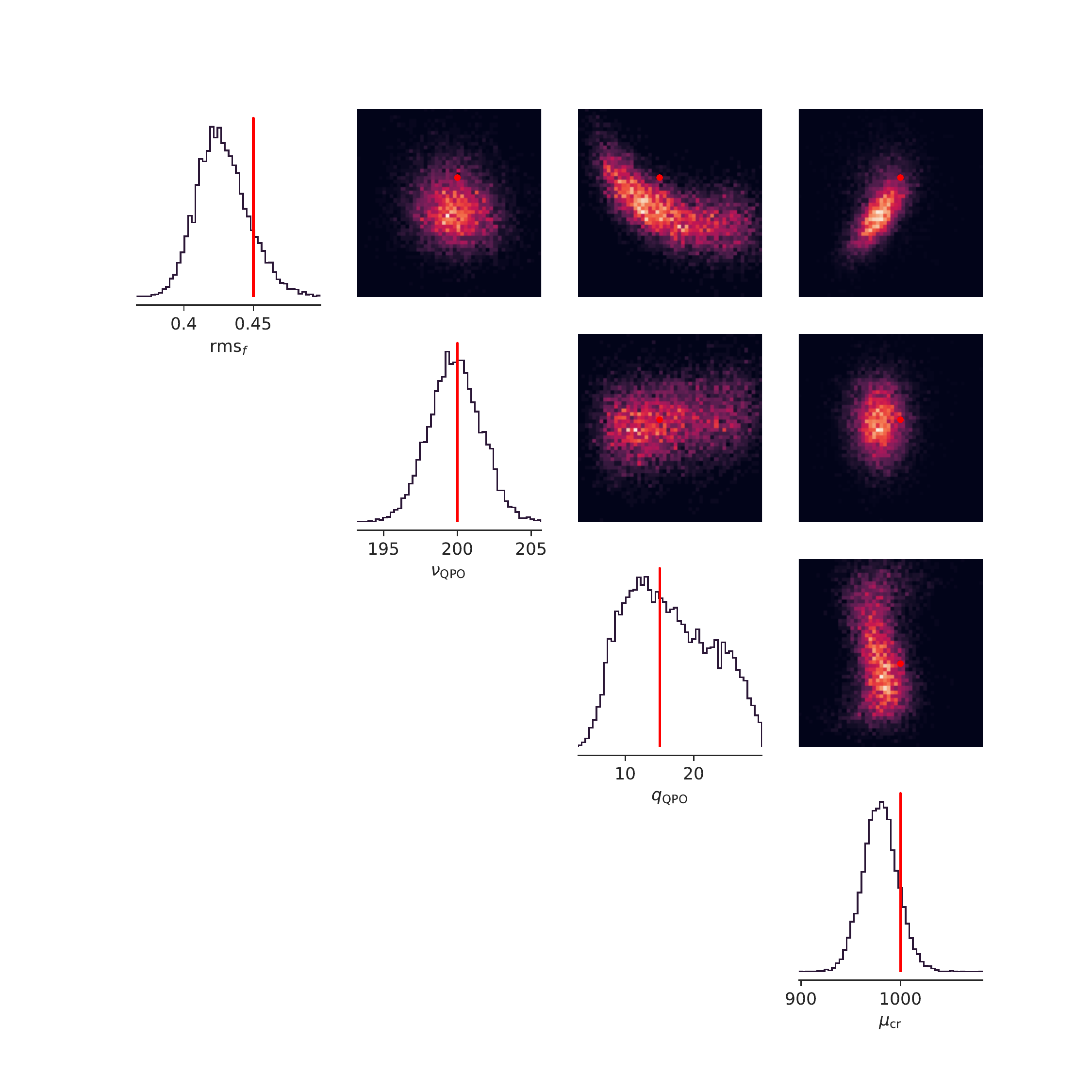}
\caption{Posterior distribution as derived through amortized SNPE: on the diagonal, we show one-dimensional marginalized posterior densities, on the off-diagonal a heat map of parameter pairs. All distributions are normalized so that they integrate to one. In red, we mark the true parameters that generated the data. For all parameters except the quality factor, the posterior clusters tightly around the true value.}
\label{fig:hf_corner}
\end{center}
\end{figure}

\begin{figure}
\begin{center}
\includegraphics[width=0.5\textwidth]{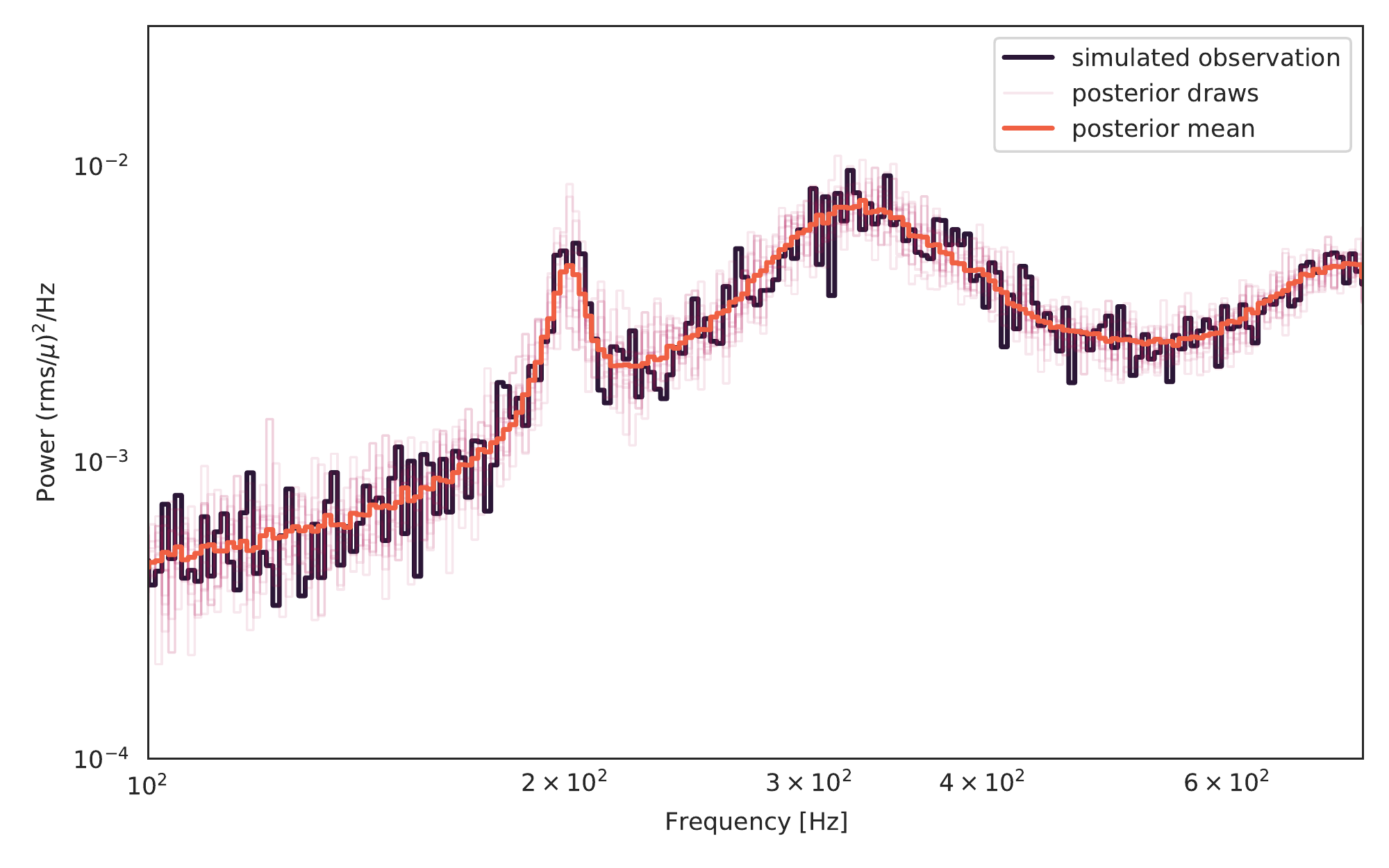}
\caption{We show the simulated observation of a QPO at $200\mathrm{Hz}$ (dark purple), along with 100 random draws from the posterior (light pink), as well as the posterior median derived from these 100 random draws (orange). The draws from the posterior clearly trace out the QPO. In addition, the posterior median makes frequency-dependent changes in the white noise level due to dead time evident.}
\label{fig:hf_post}
\end{center}
\end{figure}

\subsection{Single Periodogram: High-Frequency (HF) QPO}
\label{sec:single_hf}

While at low frequencies, the dead time effects are appreciable and strongly affect inferences e.g.~of the \rms amplitude, arguably the strongest effects due to the specific time scale that dead time imposes on the time series are at high frequencies. In particular, the periodogram is known to be ``wavy'' with saddle points at $n / 2\deadtime$ Hz. Thus, we also generate data that includes a high-frequency QPO at $\nu_0 = 200\mathrm{Hz}$ (the first saddle point) with a quality factor of $q=15$ and a fractional \rms amplitude of $\mathrm{rms}_f = 0.45$. We again use the process from \citet{timmer1995} to generate a light curve of $T = 10 \mathrm{s}$ duration with a mean incident count rate of $\mu_{cr} = 1000$, transform it into a list of events by randomizing the arrival time inside a given light curve bin, and apply dead time to the resulting event lists. We then sum the individual light curves generated to represent data from each of \nustar's individual detector modules, and calculate a periodogram with a Nyquist frequency of $\nu_\mathrm{Nyquist} = 750 \mathrm{Hz}$. The resulting light curves and periodograms for both the dead time-affected data and the data without dead time are shown in Figure \ref{fig:hf_data}.

The effect of dead time becomes much more apparent than in the periodogram for the low-frequency QPO: at high frequencies, dead time imposes a strong oscillatory structure onto the periodogram, which in turn reduces the signal-to-noise ratio of the QPO intrinsic in the simulated data. We repeat the same inference process as for the low-frequency QPO, but adjust the prior for the centroid frequency to reflect our change in expectation about the frequency of the QPO (see Table \ref{tab:priors}). We draw 50,000 parameter sets from the prior and generated an associated dead time-affected periodograms to use for training the MAF. We first again train the model on the periodogram itself, but here choose a logarithmically binned periodogram to reduce noise effects at high frequencies. We present results in Figure \ref{fig:hf_corner} and \ref{fig:hf_post}. As in the low-frequency case, the trained model successfully recovers the true parameters, though we the distributions for the fractional \rms amplitude and quality factor are wider than for the low-frequency QPO. Figure \ref{fig:hf_post} shows once again posterior draws along with a posterior median derived from $100$ simulated data sets. Overall, the SBI model recovers the shape of the QPO well, and also provides an adequate model for the dead time affecting the observations.

\begin{figure*}
\begin{center}
\includegraphics[width=\textwidth]{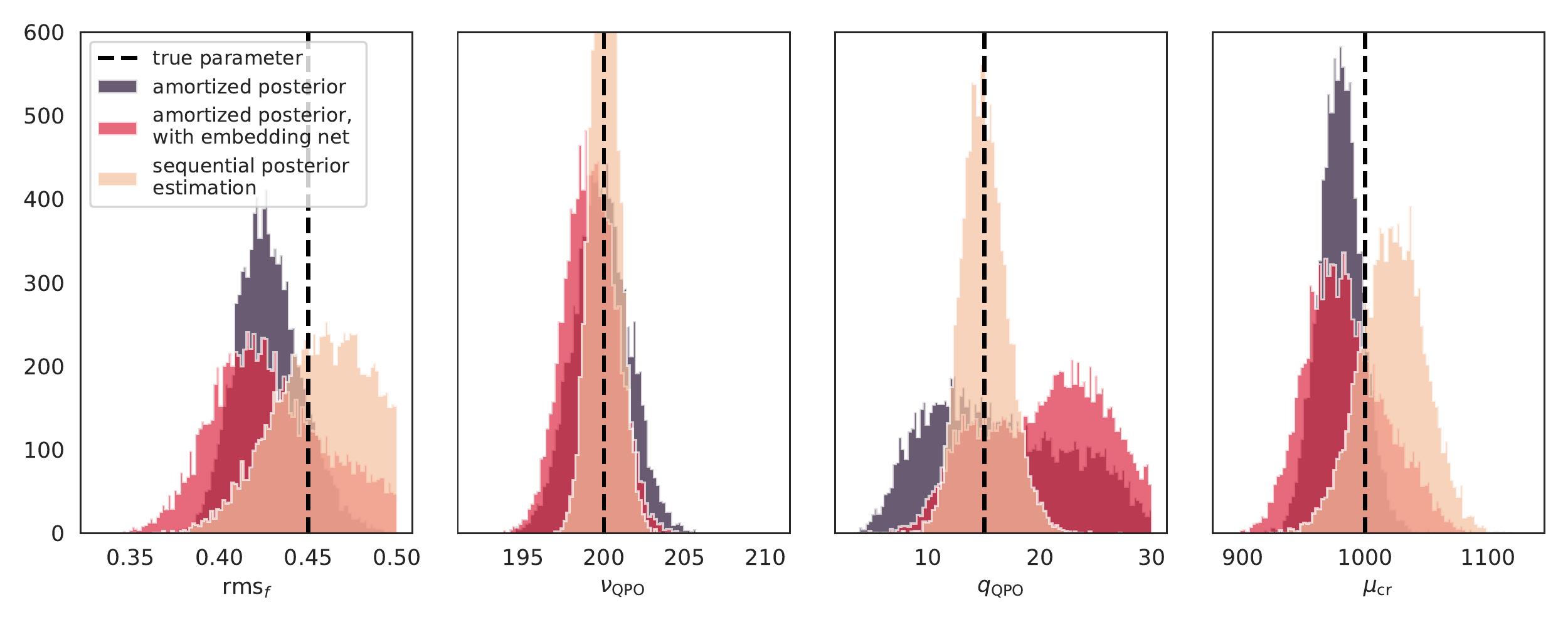}
\caption{Posterior probability distributions for three different SBI models: (1) amortized posterior inference with 50,000 simulations on the raw periodogram (purple), (2) amortized posterior inference with 50,000 simulations, with an embedding net generating summaries of the data (pink), (3) sequential posterior estimation on the raw periodogram, for five rounds of 1,000 simulations each (orange).}
\label{fig:hf_post_comp}
\end{center}
\end{figure*}

Again, we build an embedding net with a convolutional layer, a maxpooling operation, a fully connected layer, and rectified linear activation units, but explore different architectures and hyperparameters (e.g.~convolutional kernel sizes) within these bounds. As with the low-frequency QPO, we recover at most comparable performance to using the logarithmically-binned periodogram directly (see also Figure \ref{fig:hf_post_comp}). In fact, in all architectures, the embedding net favours higher quality factors compared to the model without embedding net. It is worth noting, however, that models drawn from the posterior, and the median of these draws, are virtually indistinguishable from the previous model and reproduce the observed powers almost exactly within the uncertainties. This could indicate that there is an intrinsic uncertainty in the quality factor for this QPO. Given that dead time produces excess power at these frequencies, it is not unexpected that information about the wings of the Lorentzian that produced the data are lost, and that this loss of information directly translates into a loss of precision in our inferences of the width of the QPO. 

We compare the two amortized models to posteriors produced by sequential neural posterior estimation of five rounds of 1000 simulations each, set up in the same way as for the low-frequency QPO. This approach produces well-constrained posterior distributions that are narrower than for the amortized versions, with only $5000$ simulations. A comparison of all three approaches is shown in Figure \ref{fig:hf_post_comp}. Again, we use the distance between the posterior mean and the true parameter value, expressed in standard deviations of the posterior, as a rough measure of how well the posterior captures the true underlying parameter value. As before, the sequential algorithm produces posteriors with means nearest to the true value. The largest distances between true parameter value and posterior mean  are of the order of $\sim1\sigma$, which we consider an acceptable performance (see also Table \ref{tab:hf_post_distance}).

\begin{figure}
\begin{center}
\includegraphics[width=0.5\textwidth]{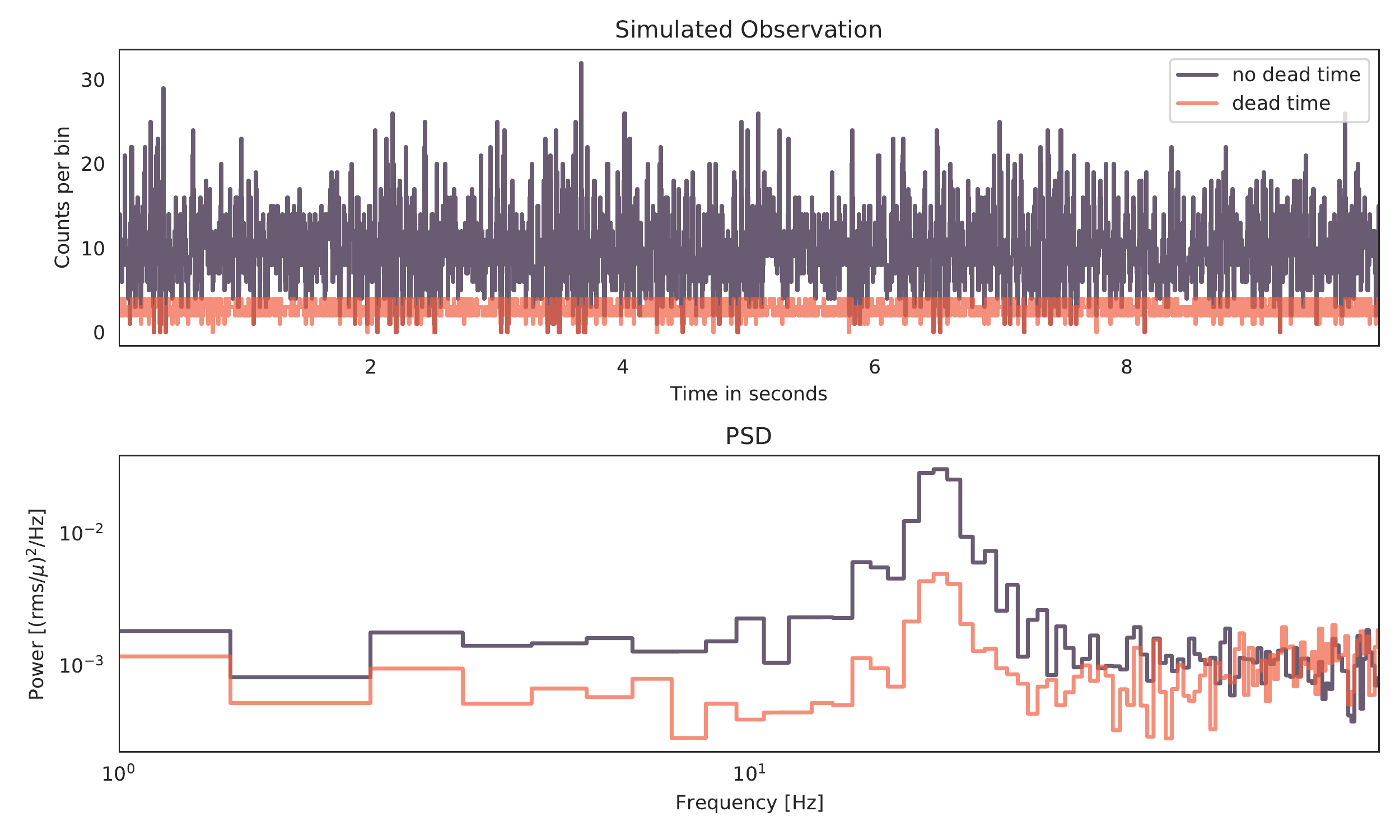}
\caption{Simulated light curve and Fourier products for a single QPO at $20\mathrm{Hz}$. Top panel: simulated light curves both with (orange) and without (purple) dead time applied. Bottom panel: averaged periodograms out of 10 $1$-second segments corresponding to the light curves in the top panel.}
\label{fig:lf_avg_sim}
\end{center}
\end{figure}

\begin{figure}
\begin{center}
\includegraphics[width=0.5\textwidth]{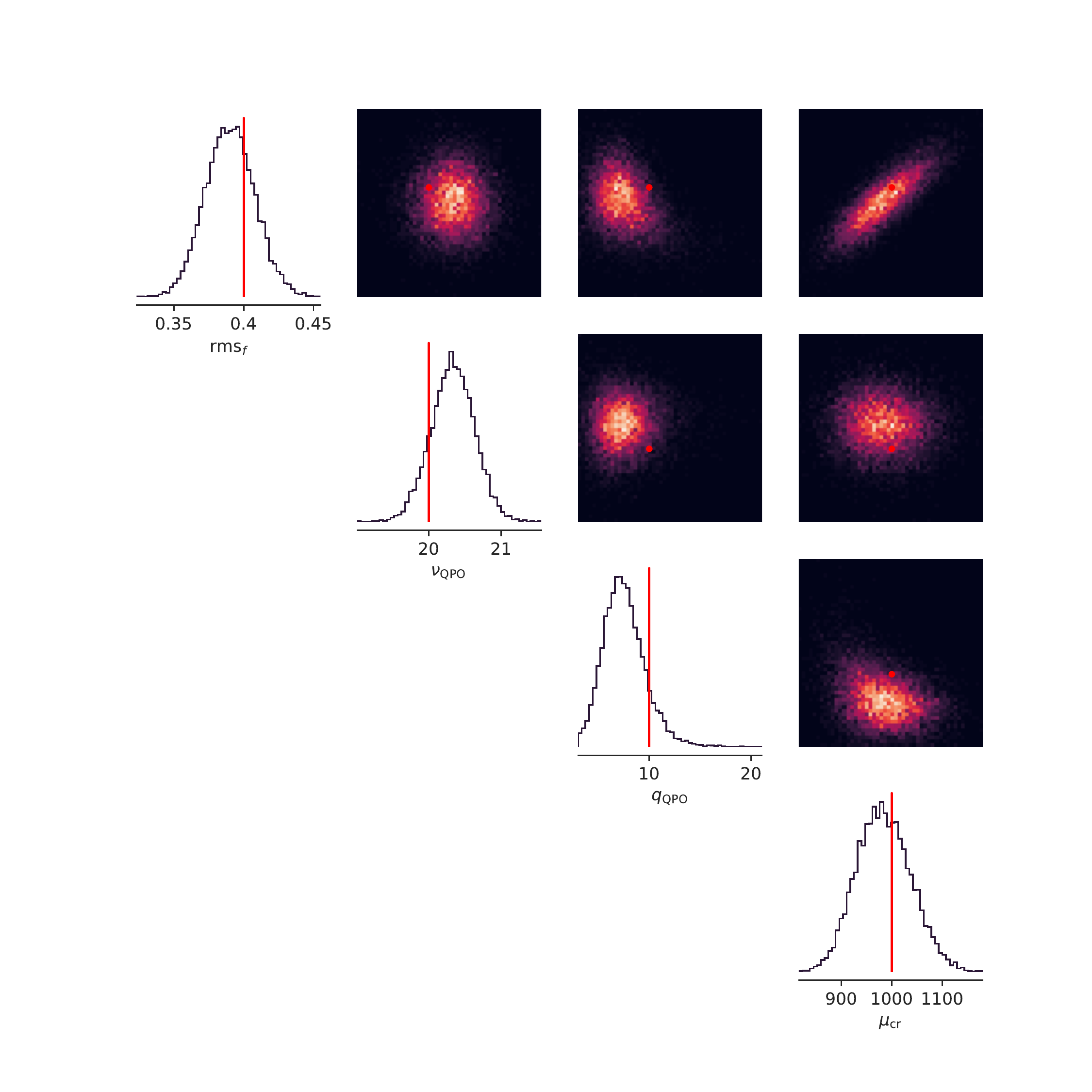}
\caption{Posterior distribution as derived through amortized SNPE for the $20\mathrm{Hz}$ QPO in an averaged periodogram: on the diagonal, we show one-dimensional marginalized posterior densities, on the off-diagonal a heat map of parameter pairs. All distributions are normalized so that they integrate to one. In red, we mark the true parameters that generated the data.}
\label{fig:lf_avg_corner}
\end{center}
\end{figure}

\begin{figure}
\begin{center}
\includegraphics[width=0.45\textwidth]{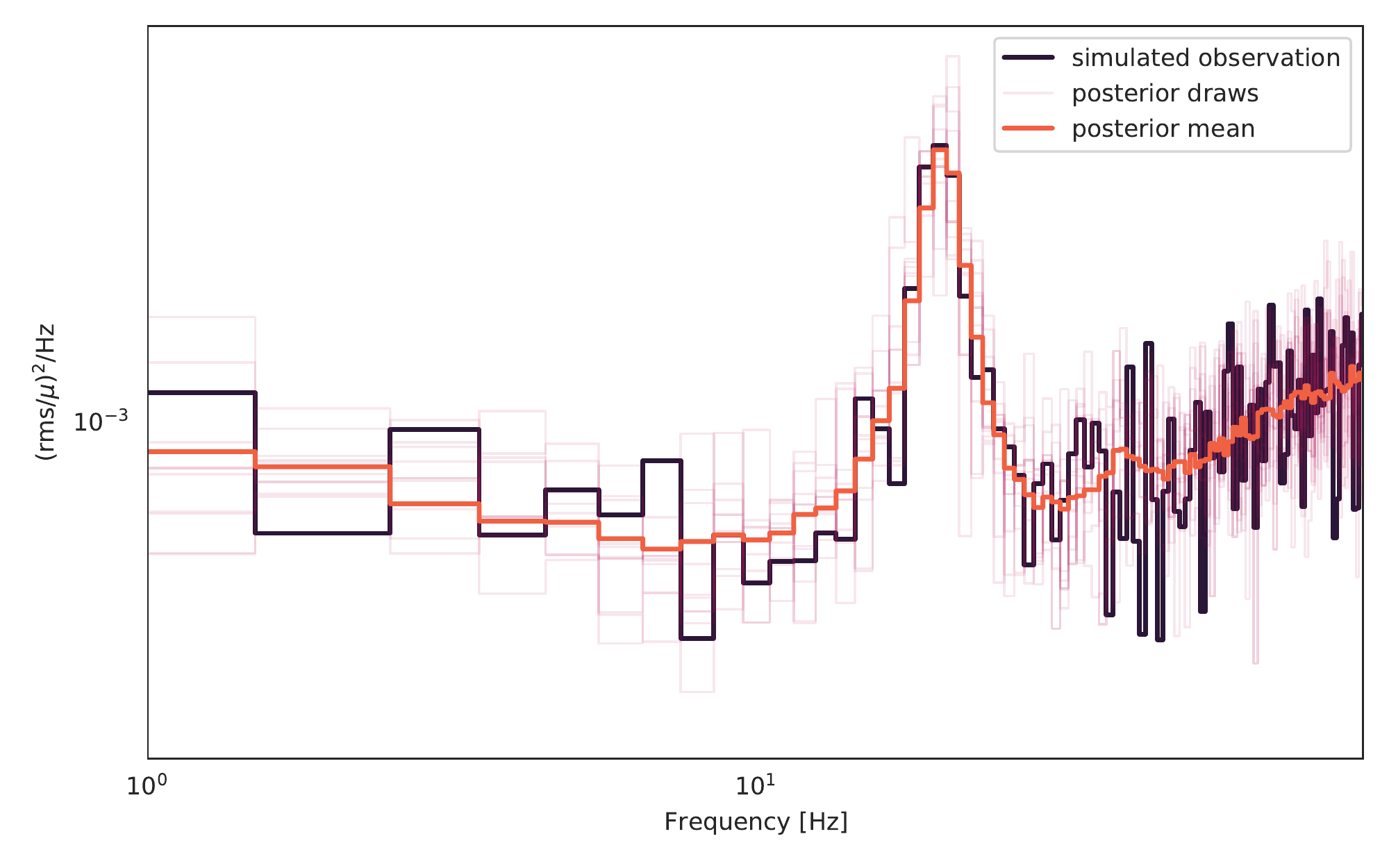}
\caption{For a low-frequency QPO at $20\mathrm{Hz}$ in an averaged periodogram, we show the simulated observation (dark purple), along with 100 random draws from the posterior (light pink), as well as the posterior median derived from these 100 random draws (orange). The draws from the posterior clearly trace out the QPO. In addition, the posterior median makes frequency-dependent changes in the white noise level due to dead time evident.}
\label{fig:lf_avg_post}
\end{center}
\end{figure}

\begin{figure}
\begin{center}
\includegraphics[width=0.5\textwidth]{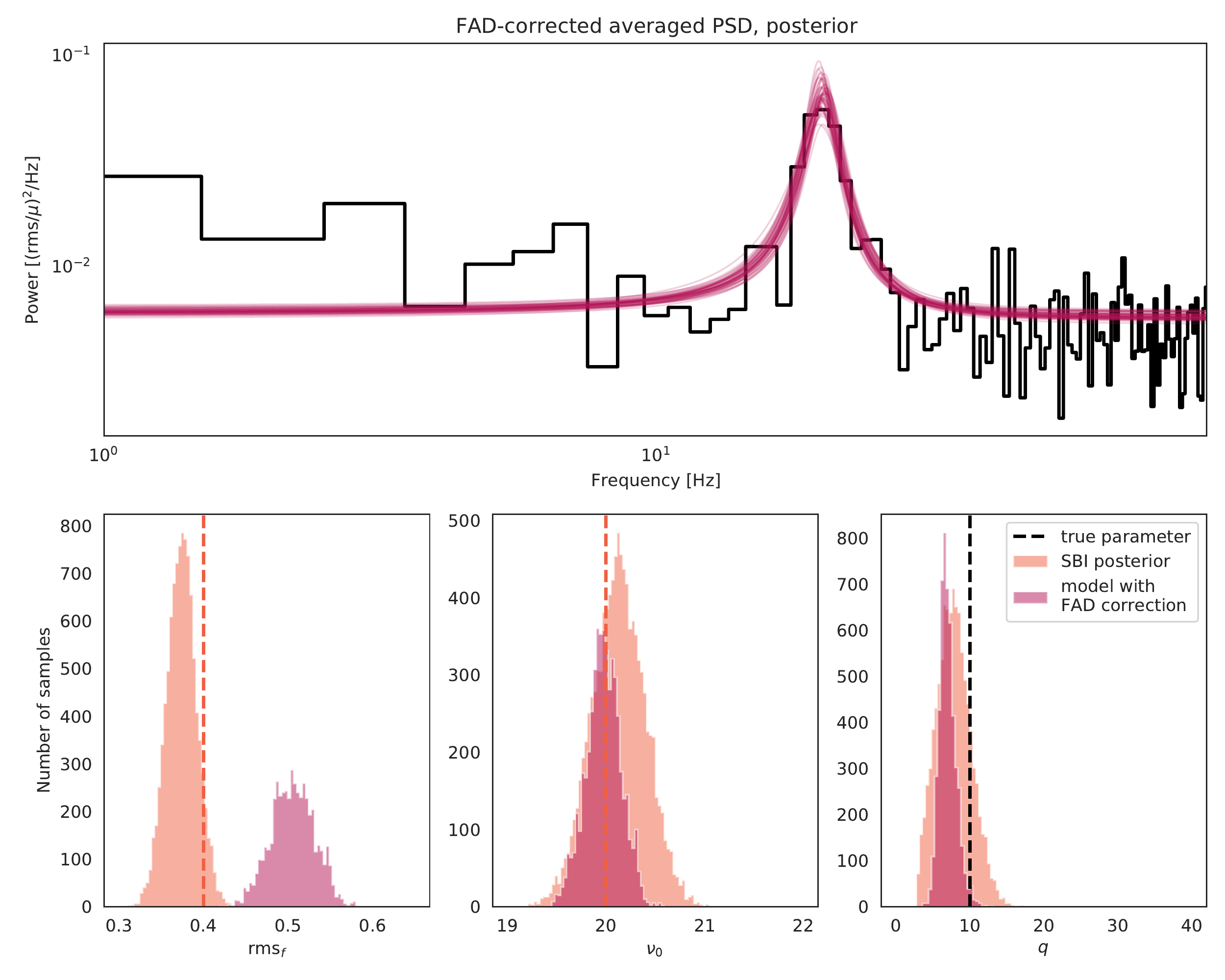}
\caption{Top: FAD-corrected periodogram with draws from the posterior probability density of a model with a $\chi^2$ likelihood and samples generated with MCMC. Bottom: Comparisons between the SNPE model and the likelihood-based model of the FAD-corrected periodogram for the three most important QPO parameters: the fractional \rms amplitude (left), the QPO centroid frequency (middle), and the QPO quality factor (right).}
\label{fig:lf_avg_sbi_fad}
\end{center}
\end{figure}

\begin{table}
\centering{
\renewcommand{\arraystretch}{1.3}
\footnotesize
\caption{HF QPO: Distance between true parameters and posterior mean}
\begin{threeparttable} 
\begin{tabularx}{8.0cm}{p{1.5cm}p{1.0cm}p{1.0cm}p{1.0cm}p{1.0cm}}
\toprule
\bf{Parameter} & \bf{True value} & $\Delta_\mathrm{am}$ & $\Delta_\mathrm{SE}$ & $\Delta_\mathrm{EN}$\\ \midrule
 $\mathrm{rms}_f$   & $0.45$     & $1.18\sigma$ & $0.27\sigma$ &  $1.17\sigma$ \\
 $\Delta\nu_0$      & $200.0$    & $0.09\sigma$ & $0.06\sigma$  &  $0.51\sigma$ \\
 $q_\mathrm{QPO}$   & $15.0$    & $0.23\sigma$ & $0.03\sigma$  &  $0.85\sigma$ \\
 $\mu_\mathrm{cr}$  & $1000.0$  & $1.13\sigma$ & $0.85\sigma$ &  $0.91\sigma$ \\
\bottomrule
\end{tabularx}
   \begin{tablenotes}
      \item{We provide a rough quantification of how well the posterior distributions recover the true parameter values through calculating the distance between the true parameters and the posterior mean, in units of standard deviations from the mean $\sigma$, for the three models (AN: amortized model; SE: sequential model; EN: model with embedding net).}
\end{tablenotes}
\end{threeparttable}
\label{tab:hf_post_distance}}
\end{table}

\subsection{Averaged Periodogram: Low-Frequency QPO}
\label{sec:avg_lf}

\citet{huppenkothen2018} showed that above $\sim 30$ averaged cospectra, the distribution of cospectral powers become approximately Gaussian. In this case, a standard Gaussian likelihood will provide a reasonably good approximation and a more traditional Bayesian modelling procedure should yield adequate results. For observations that are too short to admit averaging 30 or more segments, however, the distribution of cospectral powers is not known, and SBI might provide a viable alternative. We test its use on averaged periodograms of simulated observations analogously to Sections \ref{sec:single_lf} and \ref{sec:single_hf} for both a low-frequency QPO and a high-frequency QPO.

We used the same simulated data as in Section \ref{sec:single_lf} with a single QPO at $\nu_0=20 \mathrm{Hz}$ in an observation of duration $T=10\mathrm{s}$. Instead of Fourier-transforming the entire light curve, we subdivided the light curve into $10$ segments of $T_\mathrm{seg}=1\mathrm{s}$ duration, constructed the periodogram of each, and then averaged all ten individual periodograms. The resulting periodogram for the observation, both with and without dead time applied to the photon arrival times, is shown in Figure \ref{fig:lf_avg_sim}.

As in Section \ref{sec:single_lf}, we build three models using SBI: (1) an amortized model using $50000$ simulation generated using the same process that generated the data, (2) a sequential version with five round of $5000$ simulations each, and (3) a model that includes an embedding net to generate informative summaries of the data. In Figures \ref{fig:lf_avg_corner} and \ref{fig:lf_avg_post}, we show the results for the amortized model generated from $50000$ simulations as an example. As with the single periodogram, the model recovers the input parameters well. The sequential model generates comparable results, except for the centroid frequency, where it produces a narrower posterior around the true value, but requires only a tenth of the simulations as the amortized model. As with previous models, the embedding net does not improve the quality of the inferences, nor does it lead to computational gains. 

To understand the performance of an SBI-based model compared to alternative treatments of dead time, we compare the model to a likelihood-based model using a FAD-corrected version of the averaged periodogram. Because our \nustar-like simulations generate light curves for two independent detectors, we followed \citet{bachetti2018} and the differences in Fourier amplitudes to estimate the imprint of dead time on the periodogram. We subtracted a spline-fit model of the dead time as an approximation, and subsequently defined a model analogous to that in Section \ref{sec:single_lf}, using the same model structure and priors. We sampled this model using MCMC and present posterior draws and the FAD-corrected periodogram in Figure \ref{fig:lf_avg_sbi_fad}. We find that overall, the model represents the FAD-corrected periodogram well. Comparing the distributions derived from both SNPE and the likelihood-based model (Figure \ref{fig:lf_avg_sbi_fad}), the latter overestimates the fractional \rms amplitude (as already observed by \citealt{bachetti2018}), but provides broadly similar performance for the centroid frequency and the quality factor. The posterior derived from SNPE are somewhat wider for the latter two parameters; that is also expected given that it is derived from noisy simulations. 

\subsection{Averaged periodogram: high-frequency QPO}
\label{sec:avg_hf}

We repeat the process of the previous section, but with a high-frequency QPO at $\nu_0 = 400 \mathrm{Hz}$. This frequency is higher than the QPO simulated in Section \ref{sec:single_hf}, an intentional choice to allow us to prove the response of the model to a QPO in a different part of the highly dead-time affected part of the periodogram. We use SNPE on both the raw periodogram powers, as well as a logarithmically binned version of the periodogram, and find consistently better inferences for the latter. This is not surprising, as the additional averaging dampens some of the noise variance at high frequencies. 

As in the rest of this Section, we run the same three SNPE models, and find overall good results with all models. We find that all three produce acceptable posteriors, with the amortized model once more producing the best-constrained distributions for the smallest computational cost. 

\begin{figure*}
\begin{center}
\includegraphics[width=\textwidth]{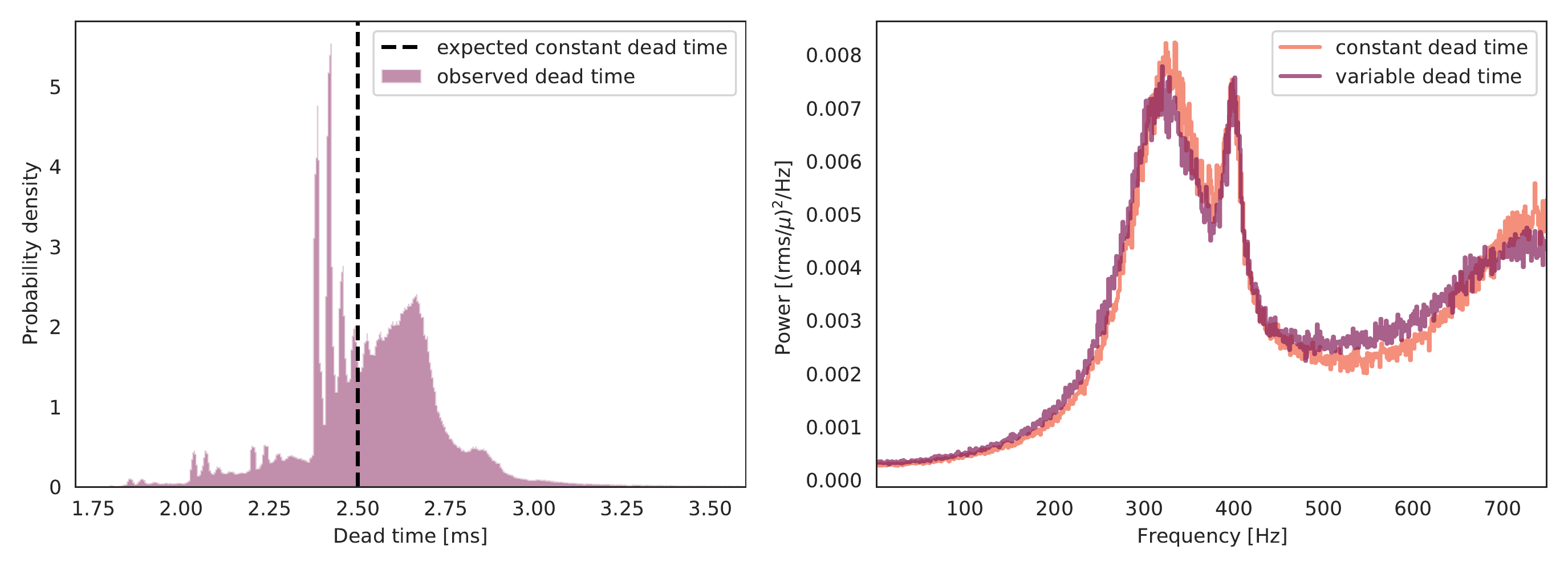}
\caption{Left panel: a histogram of dead time intervals derived from the \nustar~observation \texttt{80401312002} reveals a highly structured distribution around the nominal constant value of $0.0025\mathrm{s}$. Right panel: the effect of variable dead time on the periodogram becomes apparent in an averaged periodogram consisting of $500$ segments of $1\mathrm{s}$ duration: at high frequencies, the periodogram derived with variable dead time (pink) deviates significantly in shape from the periodogram generated from data assuming constant dead time (orange).}
\label{fig:nustar_deadtime}
\end{center}
\end{figure*}

\subsection{Variable Dead Time}
\label{sec:variable}

Dead time is related to the amount of time the flight software takes to process a given event. This could be related to the properties of the single event, e.g., how many pixels are involved in the detection (event ``grade''), but also on the subtleties of the flight software itself.
Therefore, dead time can change considerably between different events, departing from the assumption of constant dead time we have used so far.
Indeed, the dead time for \nustar~is not constant: it varies slightly around an assumed value of $\deadtime = 0.0025\mathrm{s}$. 
However, most dead time models--including the simulator we use above as part of the SNPE model--assume that a constant dead time of $\deadtime$ is a reasonable approximation. In principle, the SBI method introduced here can take variable dead time into account: as long as we have some understanding of the underlying process that generated the dead time and can implement it in the simulator that generated the data, it can be part of the model, at the potential loss of computational efficiency due to the additional operations introduced as part of letting $\deadtime$ vary. We implement a simulator with variable dead time to (1) test whether a model that assumes constant dead time generates biased results when dead time is not, in fact constant, and (2) whether taking variable dead time into account in the SBI model produces significantly better inferences. 

We constructed an empirical distribution of dead times for \nustar~from the observation of GRS 1915+105 introduced in Section \ref{sec:dataprocessing}. We calculated the intervals between photon arrival times for all $N=3572101$ photons recorded in both detector modules, and corrected these intervals for the live time of the detector since the last event recorded in the column \texttt{PRIOR}. The remainder constitutes the dead time caused by the previous event.
Because some non-scientific events like shield vetoes reset the live time, but the associated events are not recorded, we expect a small number of outliers in the distribution.

In Figure \ref{fig:nustar_deadtime}, we present a histogram of the dead time for this observation: it is apparent that the dead time values follow a complex, multi-modal distribution around the generally accepted value of $\deadtime=0.0025\mathrm{s}$. This distribution was corrected for outliers caused by the aforementioned shield vetoes after an initial visual inspection. To be conservative, we only removed dead time intervals larger than $10$ ms. Even so, \nustar~dead time can vary from below $2\mathrm{ms}$ to above $3\mathrm{ms}$. 

In order to explore the effect of the non-constant nature of \nustar~dead time, we built a simulator that takes it into account: instead of applying a constant dead time to each photon, and removing photons falling within that dead time, for each simulated photon we draw from the real, empirical distribution of dead times from this observation of GRS 1915+105, and remove photons that fall within the specific value of dead time for a given photon. Because standard periodograms are noisy and therefore the subtle effects of variable dead time will be difficult to discern visually, we simulate a long light curve of $500\mathrm{s}$ and construct an averaged periodogram out of 500 $1\mathrm{s}$ segments, in order to reduce the noise in the periodogram enough to illustrate the differences between variable and constant dead time. We include the high-frequency QPO from Section \ref{sec:avg_hf}, since dead time depends on total flux, and we expect the sources we observe to be highly variable in practice. In Figure \ref{fig:nustar_deadtime} (right panel), we show the periodogram of a light curve affected with variable dead time, as well as the periodogram corresponding to the same list of events, but filtered with a constant dead time of $\deadtime = 0.0025\mathrm{s}$. There are appreciable deviations in the shape of the dead time-affected power spectrum when dead time is not constant, though it is unclear from this plot alone how different the resulting posterior probability distributions are.

To test this, we simulated the same photon arrival times as in Section \ref{sec:avg_hf}, but used the procedure above where the dead time is not constant, but randomly drawn from the empirical distribution derived from the observed \nustar~data. We generated a $10\mathrm{s}$ light curve, which was used to produce an averaged periodogram out of 10 $1\mathrm{s}$ segments. 

\begin{figure*}
\begin{center}
\includegraphics[width=\textwidth]{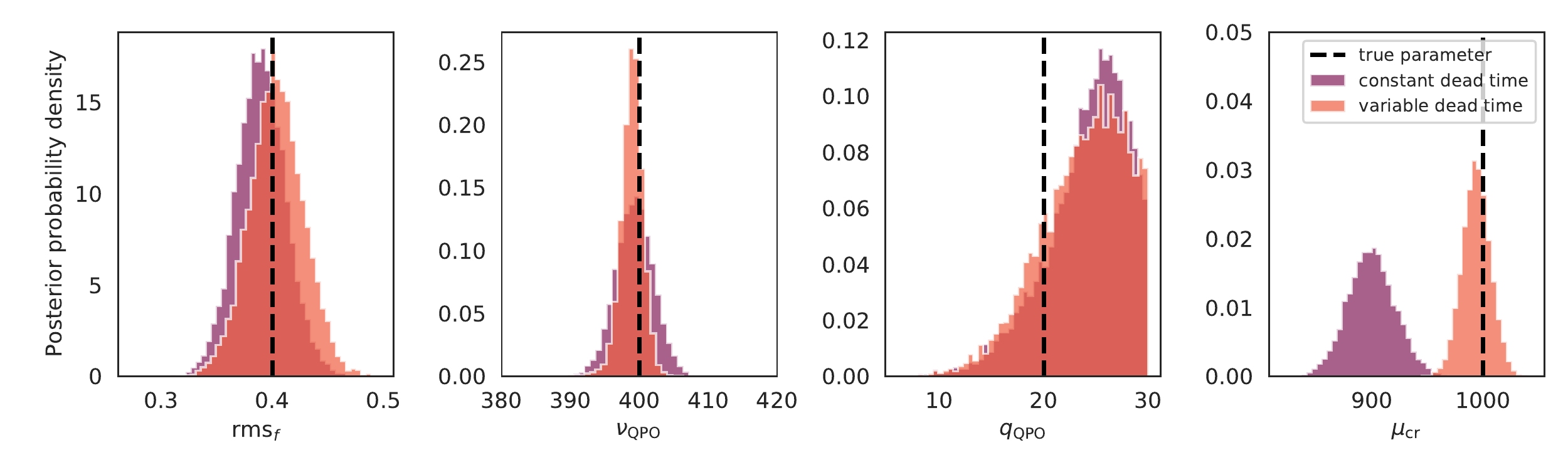}
\caption{Marginalized posterior probability distributions for simulated data with variable dead time, using a model with constant dead time (purple) and variable dead time (orange) to infer QPO and light curve parameters. The constant dead time model compensates for the underestimation of the true dead time by inferring a lower incident count rate, whereas the variable dead time model accurately recovers all four parameters.}
\label{fig:nustar_comparison}
\end{center}
\end{figure*}

We then used SNPE to draw from the posterior, using the simulator with the \textit{constant} dead time representation. This aims to reproduce the situation we expect to see in real applications: observations affected by variable dead time, modelled with a simulator that assumes dead time is constant. We use $5$ rounds of 1000 simulations each on the raw periodogram powers to allow comparison with the SNPE model from Section \ref{sec:avg_hf}. We find that the model produces posteriors that accurately recover the true input parameters (Figure \ref{fig:nustar_comparison}, purple distributions), except for the incident count rate, where the model produces a posterior that centres on $900\,\mathrm{counts}/\mathrm{s}$ and excludes the true value of $\mu_\mathrm{cr} = 1000\,\mathrm{counts}/\mathrm{s}$. This may be related to the fact that the mean dead time derived from the observation used to generate the empirical distribution is not exactly $2.5\mathrm{ms}$, but rather $2.57\mathrm{ms}$; the additional dead time can, in the limit of observing a very high count rate, account for the additional loss of photons. It appears, however, that the model efficiently compensates for this by assuming a lower incident count rate. It may be possible, in the future, to infer a better constant dead time value together with the other model parameters. In order to test whether this effect is also relevant for single short segments, we repeat the set-up above, but generate a periodogram of the full $10\mathrm{s}$ segment. Again, we use SNPE with a simulator assuming constant dead time, and discover that the same effect is also present when the periodogram is not averaged (i.e.~uncertainties are large): parameter inferences are comparable to Section \ref{sec:avg_hf} except for the incident count rate, which is underestimated.

In order to confirm our interpretation of our posteriors, we also generate posteriors based on the simulator that uses the empirical distribution. That is, the simulations used to generate the posterior distribution will also include variable dead time. For this model, we required an additional three rounds in the sequential inference process, for a total of $8000$ simulations. The posterior probability densities are comparable to those with the constant dead time model in all parameters except for the incident count rate, where this model accurately recovers the true incident count rate (Figure \ref{fig:nustar_comparison}, orange distributions). This indicates that indeed, assuming constant dead time will lead to biased inferences in the incident count rate and, thus, in the total flux. We repeat this analysis with simulated data sets that include QPOs at multiple other frequencies: at $35 \mathrm{Hz}$, $168 \mathrm{Hz}$, $250 \mathrm{Hz}$ and $450 \mathrm{Hz}$. It is in principle plausible that the bias in the inference for the incident count rate depends on QPO frequency. However, we broadly find that not to be true. For all frequencies above $100\mathrm{Hz}$, the model produces accurate, well-bounded posteriors for the QPO parameters, and a biased posterior for the incident count rate around $\mu_{\mu_\mathrm{cr}} = \sim900\,\mathrm{counts}/\mathrm{s}$ and a variance of $\sigma_{\mu_\mathrm{cr}} \sim 20\,\mathrm{counts}/\mathrm{s}$, five standard deviations away from the true value. For the QPO at $35\mathrm{Hz}$, the marginalized posterior for the incident count rate actually has a mean of $\mu_{\mu_\mathrm{cr}} = 1081\,\mathrm{counts}/\mathrm{s}$, but the distribution is also close to five times wider, with a standard deviation of $\sigma_{\mu_\mathrm{cr}} = 98\,\mathrm{counts}/\mathrm{s}$. 

This discrepancy can be explained by the information available to the model: for QPOs with centroid frequencies above 100 Hz, we generally used periodograms that are sampled up to a Nyquist frequency of $\nu_\mathrm{Nyquist} = 500\mathrm{Hz}$. For a low-frequency QPO like the one at $35\mathrm{Hz}$, sampling that high would be computationally wasteful in practical applications, thus we only sampled up to $\nu_\mathrm{Nyquist}=100\mathrm{Hz}$. Consequently, the model has much less information available on the dead time affecting the light curve if the periodogram cuts off at $100\mathrm{Hz}$. As a result, the posterior distribution reflects the higher uncertainty in the incident count rate in the much larger standard deviation. 

If accurate inference of the incident count rate is required and the instrument's time resolution allows for it, we recommend generating a periodogram that includes the frequency range where dead time effects are prominent, and employing the variable dead time model. 
However, the variable dead time model adds significant computational overhead onto the calculations by a factor of $\sim 2$. In cases where inferences do not rest critically on the incident flux, using a constant dead time model may be computationally more expedient. If the incident flux is important to the physical inferences to be made, then one may implement an empirical dead time model as suggested above in order to accurately treat variable dead time.

\begin{figure*}
\begin{center}
\includegraphics[width=\textwidth]{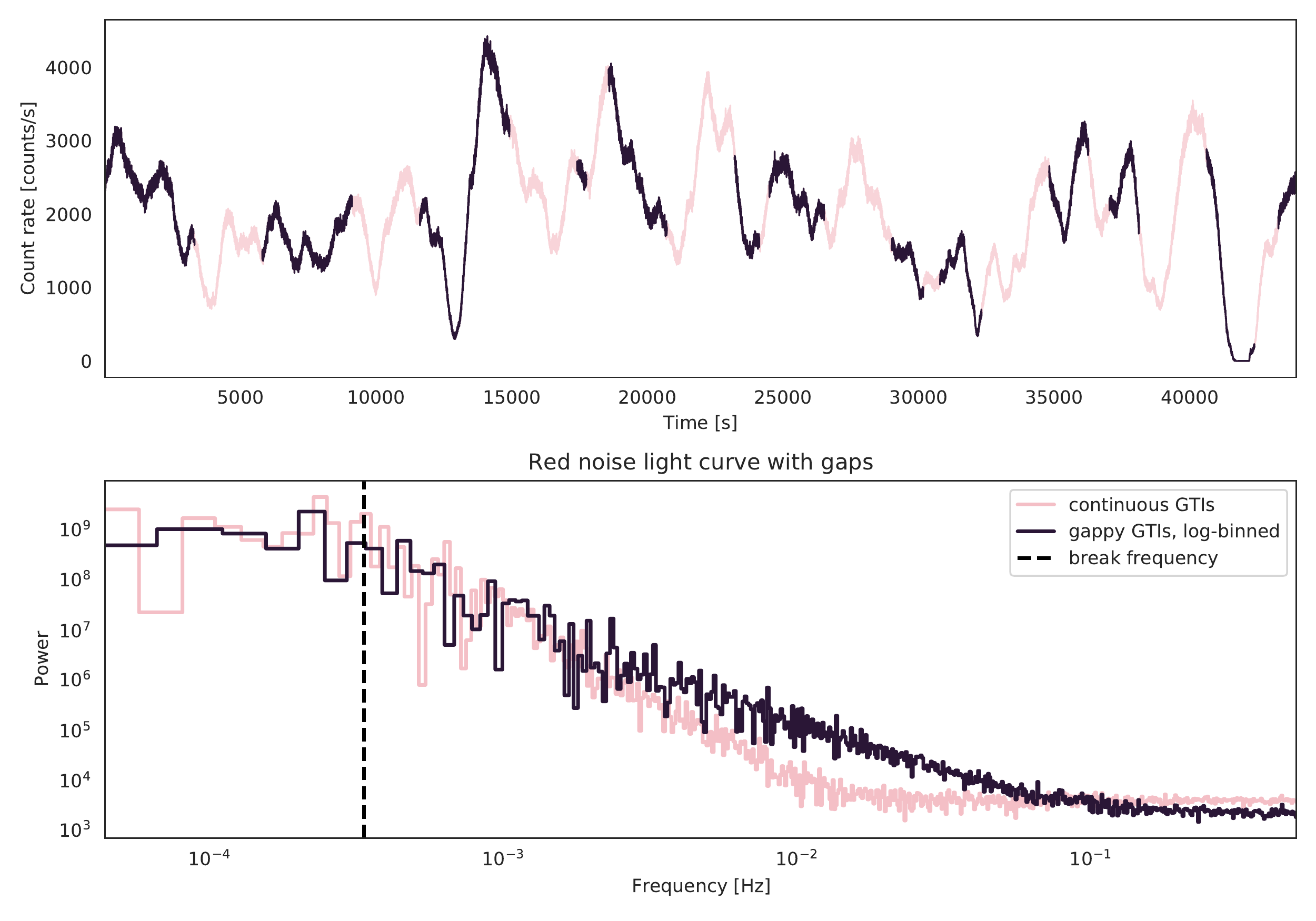}
\caption{Top: Light curve simulated from a broken power law power spectrum. In dark purple, the observed segments. Light pink marks the gaps and shows the data points that would have been observed if these gaps did not exist. Bottom: The logarithmically binned periodogram for the data without the GTIs imposed (light pink), and with the GTIs imposed (purple). The imprint of the gaps is most strongly visible in the high-frequency power law index, where the gaps push the index towards a smaller value. The dashed line marks the location of the break frequency.}
\label{fig:rn_data}
\end{center}
\end{figure*}

\begin{table*}
\renewcommand{\arraystretch}{1.3}
\footnotesize
\caption{Priors used in the models}
\begin{threeparttable} 
    \begin{tabularx}{18cm}{p{6.5cm}p{2.5cm}p{4.5cm}p{4.5cm}}
        \toprule
         \bf{Model} & \bf{Parameter} & \bf{Meaning} & \bf{Probability Distribution} \\ \midrule
        {\it red noise SBI model} & $\mathrm{rms}_f$ & fractional \rms amplitude & $\mathrm{Uniform}(0.1,0.5)$ \\
        & $\alpha_1$ & power law index below $\nu_\mathrm{break}$ & $\mathrm{Uniform}(-0.2,5.0)$ \\
        & $\alpha_2$ & power law index above $\nu_\mathrm{break}$ & $\mathrm{Uniform}(-0.2,5.0)$ \\
        & $\log(\nu_\mathrm{break})$ & break frequency & $\mathrm{Uniform}(\log(10^{-4}),\log(10^{-2})$ \\
        & $\mu_\mathrm{cr}$ & mean incident photon count rate & $\mathrm{Uniform}(500,1500)$ \\\midrule
        {\it red noise MCMC model}$^\emph{a}$ & $A_\mathrm{rn}$ & red noise amplitude & $\mathrm{Uniform}(10^7, 10^9)$ \\
        & $A_\mathrm{wn}$ & white noise amplitude & $\mathrm{Uniform}(10^2, 10^4)$ \\
        \bottomrule
    \end{tabularx}
    \begin{tablenotes}
        \item{An overview over the model parameters with their respective prior probability distributions for the red noise models in this section.}
        \item[\emph{a}]{The MCMC model shares $\alpha_1$, $\alpha_2$ and $\nu_\mathrm{break}$ with the SBI model; fractional \rms amplitude and mean count rate are reparametrized as two amplitude parameters.}
    \end{tablenotes}
\end{threeparttable}
\label{tab:rn_priors}
\end{table*}

\subsection{Computational Requirements}

We run all simulations and models on a machine with 2 AMD EPYC 7401 processors (48 CPU cores or 92 threads total) and 1TB RAM. The simulator itself is not parallelized, and a single simulation runs in about $0.24\mathrm{s}$ for all simulations in this section (simulations with variable dead time are slightly slower due to the additional random draws, but not substantially so). Thus, total time on a single CPU for 50,000 simulations is about 3 hours, though this process is trivially parallelizable.

The amortized models in this section are trained (without embedding nets) over $~40$ epochs in $\sim 20$ minutes when all cores are available. The addition of embedding nets (and consequently of parameters) always requires drastically more epochs for convergence (of the order of 150), and consequently also more than triple the training time. Once trained, inference with all models becomes very fast: drawing $10,000$ samples from the posterior takes only $\sim 1.5\mathrm{s}$ or less.

When training the sequential version of the algorithm, we generate 1000 simulations in each of the five training rounds, and alternate the generation of simulations with training a model. For the simulations in this section, this inference process takes $\sim 30$ minutes, though the single-core generation of simulations is the clear computational bottleneck of this procedure (about 24 of those 30 minutes are spent on generating simulations). Parallelizing the simulator would drastically improve efficiency in practical circumstances. Similarly, training the neural networks on a modern GPU would likely significantly reduce computation time for the amortized models. We find no significant computational differences between the different scenarios we test: both low-frequency and high-frequency QPO, as well as single PSD and averaged PSD cases have similar simulation/training requirements. We focus in this paper on the generation of segments that are extremely short (10s) compared to the typical lengths of observations (1ks - 100ks). Simulating full observations similar to those generated by modern, sensitive instruments like \nicer\ with millions of photons might strain both computational and memory resources without further optimization.

\subsection{Low-Frequency Variability in Light Curves with Gaps}\label{sec:agnbreaks}

Dead time is not the only instrumental effect that imposes significant biases onto the data. Most instruments cannot record continuous, unbroken data streams: ground-based instruments are constrained by the sun, by weather and observing constraints. Space-based instruments are limited by the need to process and downlink data, and natural effects interrupting data collection such as the South Atlantic Anomaly. As a result, long observations will contain gaps, in X-ray astronomy often defined by \textit{Good Time Intervals} (GTIs) describing the intervals during which the instrument yielded data appropriate for scientific studies. 

These gaps can make it difficult to study processes at time scales longer than the typical length of a GTI, because the windowing of the GTIs will impose their own imprint on a standard Fourier periodogram. Alternative approaches include the Lomb-Scargle periodogram \citep{lomb1976,scargle1982} and a recent method by \citet{wilkins2019} that employs Gaussian Process models to fill the gaps in the observations with realistic simulated data points. The simulation-based modelling approach employed here for dead time can also straightforwardly be used for studies of variability at very low frequencies in the presence of gaps. If a model for the variability exists and can be used as a simulator, we can construct simulated data sets and impose the same GTIs on these simulations as the instrument did on the real data during the collection process.

We simulate a $44\mathrm{ks}$ observation of a stochastic light curve generated using a power spectrum comprising a broken power law with a very low-frequency break at $\nu_\mathrm{break} = 0.3\mathrm{mHz}$, equivalent to a timescale of $3000\mathrm{s}$. The broken power law we use is given by 

\begin{equation} \label{eqn:bpl}
    f(\nu)= 
\begin{cases}
    A\left(\frac{\nu}{\nu_\mathrm{break}}\right)^{-\alpha_1}, & \text{if } \nu < \nu_\mathrm{break} \\
    A\left(\frac{\nu}{\nu_\mathrm{break}}\right)^{-\alpha_2}, & \text{if } \nu \geq \nu_\mathrm{break} \\
\end{cases}
\end{equation}

\noindent We use the real GTIs from the observation presented in Section \ref{sec:dataprocessing} to introduce realistic gaps into the data. Because we are interested in the behaviour of the model in the presence of gaps, no dead time was included in the simulations. The simulated data is shown in Figure \ref{fig:rn_data}. Other parameters of the model are the fractional \rms amplitude $\mathrm{rms}_f$, the low-frequency power law index $\alpha_1$, the high-frequency power law index $\alpha_2$, and the mean count rate $\mu_\mathrm{cr}$. We built a simulator that generates simulated observations using Equation \ref{eqn:bpl}, given a set of parameters, and uses the \nustar~GTIs from observation \texttt{80401312002} and use this simulator in combination with the SNPE algorithm, comparing logarithmically binned periodograms. 

\begin{figure}
\begin{center}
\includegraphics[width=0.5\textwidth]{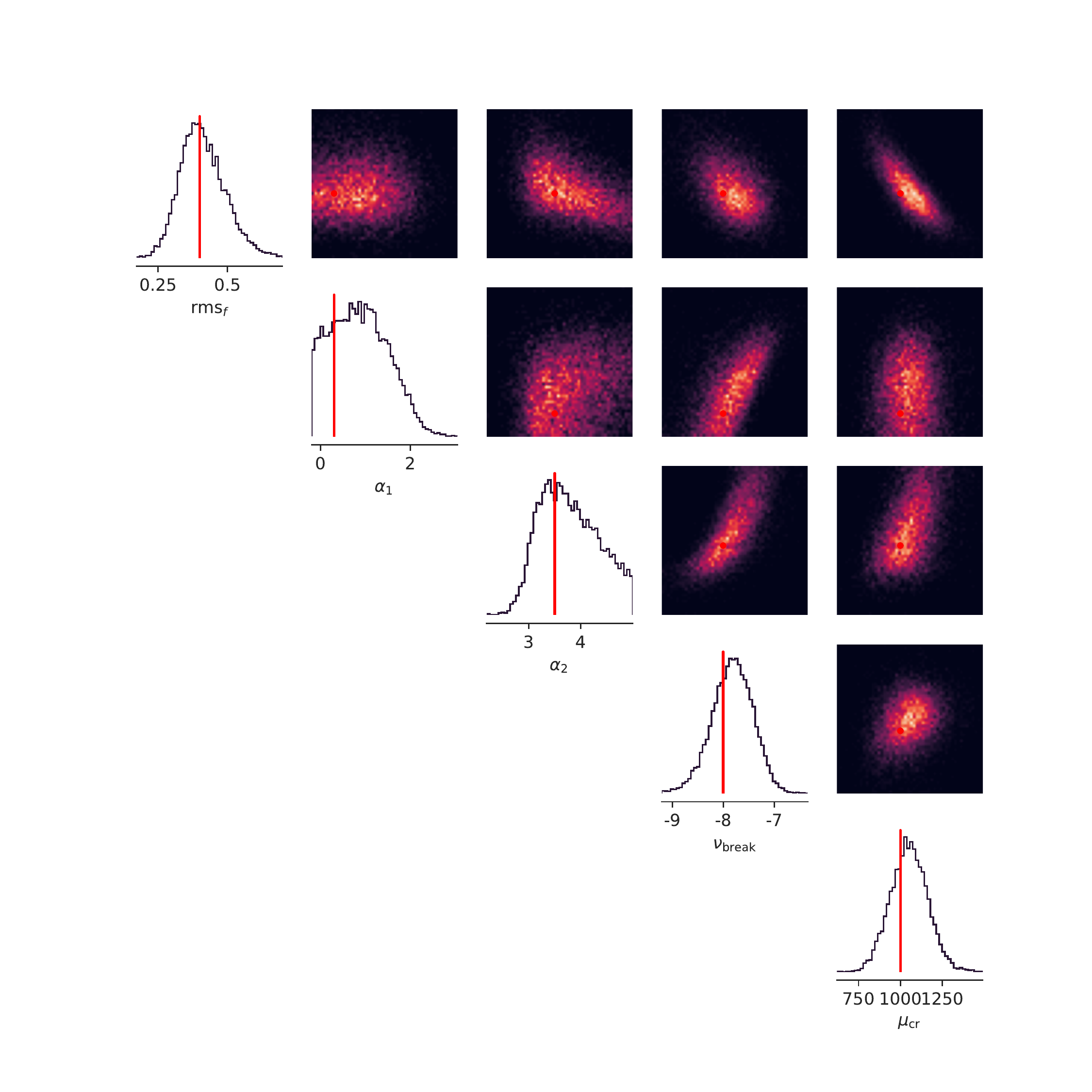}
\caption{Marginalized posterior probability distributions for the red noise model.}
\label{fig:rn_corner}
\end{center}
\end{figure}

\begin{figure}
\begin{center}
\includegraphics[width=0.5\textwidth]{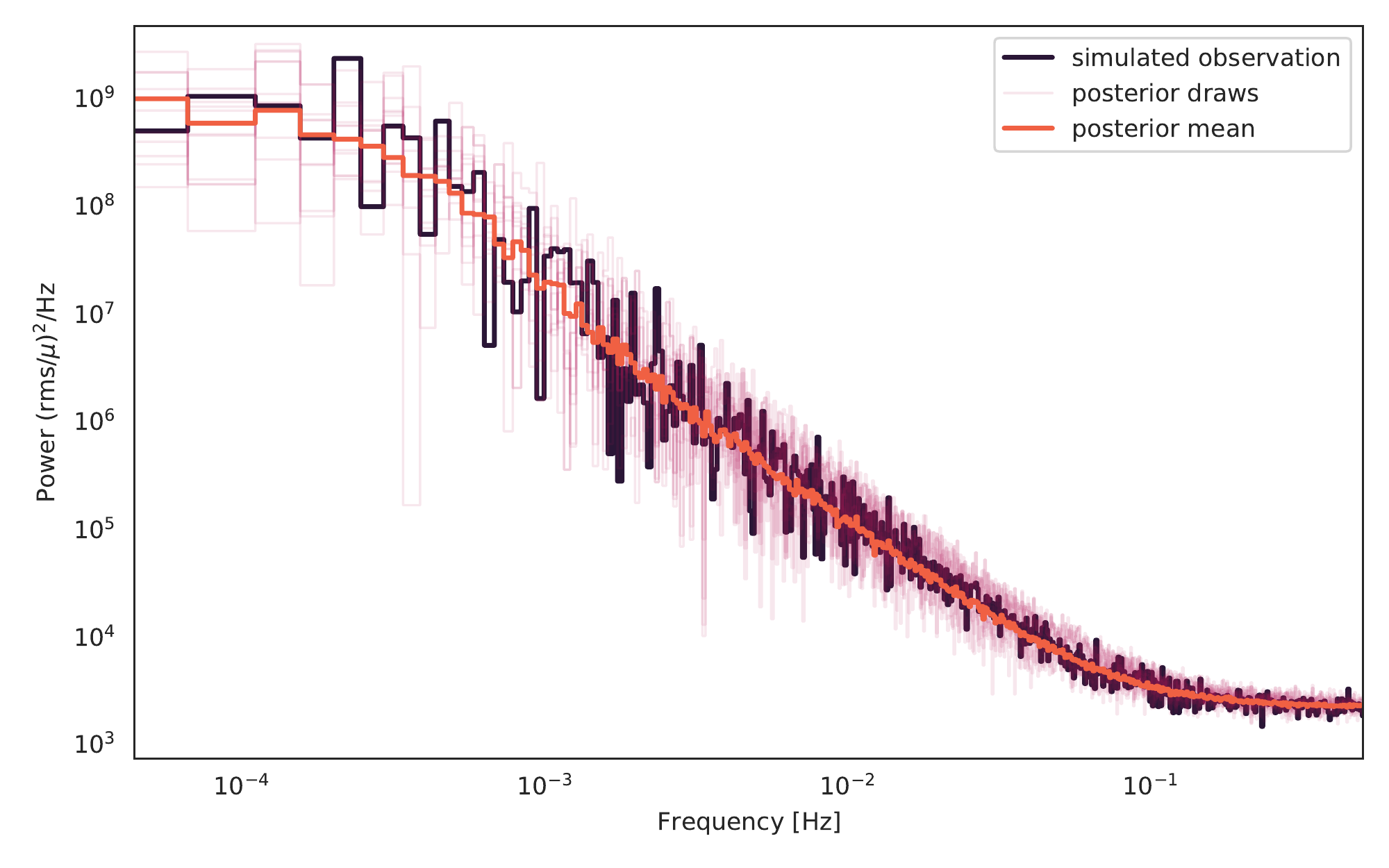}
\caption{Purple: simulated observation of a broken power law-type red noise process with gaps; pink: realizations drawn from the posterior probability distribution displayed in Figure \ref{fig:rn_corner}; orange: posterior median from 100 posterior realizations.}
\label{fig:rn_post}
\end{center}
\end{figure}

We run 10 rounds of sequential posterior estimation, each with $1000$ simulations (see Table \ref{tab:rn_priors} for the priors used in this model). The results are presented in Figures \ref{fig:rn_corner} and \ref{fig:rn_post}, respectively. The posterior probability is relatively narrow for all parameters except for the low-frequency power law index. The latter result is unsurprising, given that there are only $\sim 7$ frequency bins below the break, which naturally makes estimation of that index difficult given the typical intrinsic noise in the periodogram. More importantly, however, the break frequency is well constrained and accurately inferred, as is the high-frequency power index, though the latter has a fairly broad distribution.

\begin{figure*}
\begin{center}
\includegraphics[width=\textwidth]{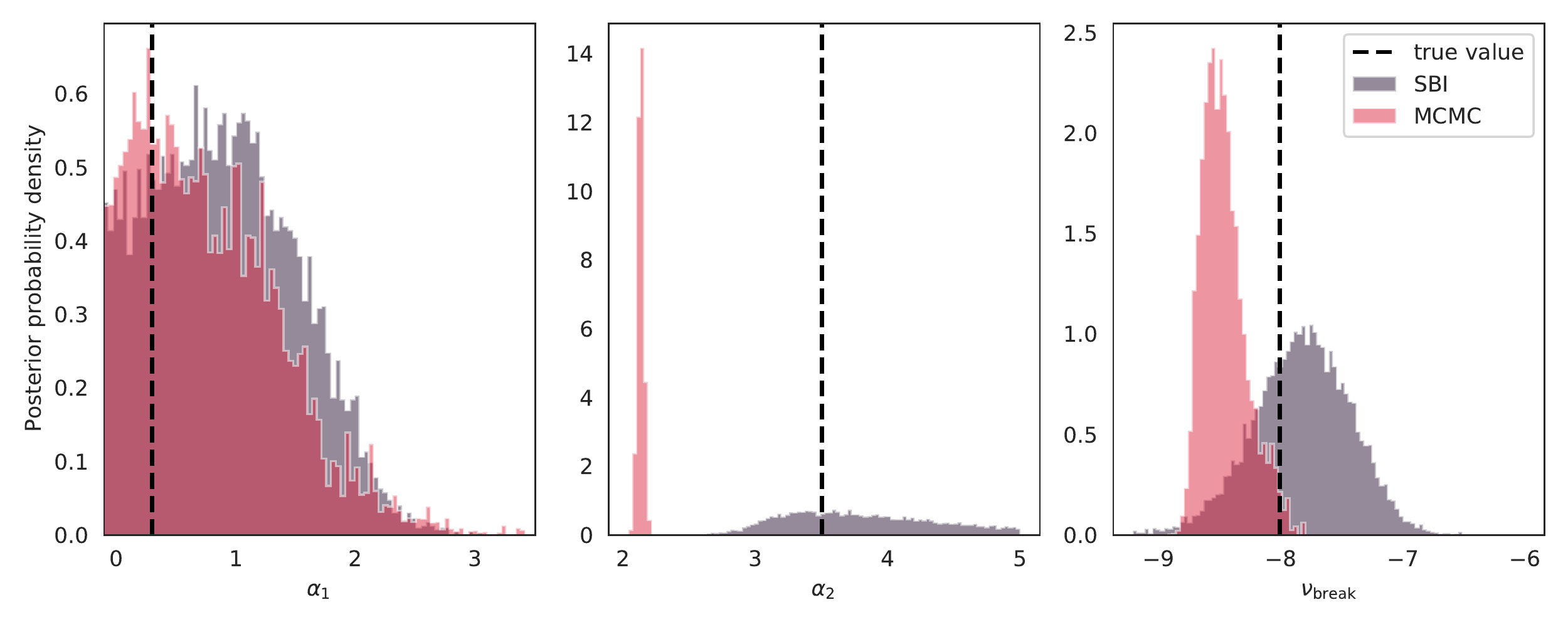}
\caption{Comparison of posterior inferences for three critical parameters in the red noise model: the power law spectral indices $\alpha_1$ (left panel) and $\alpha_2$ (middle panel) below and above the break frequency $\nu_\mathrm{break}$ (right panel), respectively. In purple, we present the marginalized posterior distributions inferred with simulation-based model and SNPE. In pink, the marginalized posterior distributions inferred using a $\chi^2$ likelihood under the assumption that there are no gaps affecting the periodogram.}
\label{fig:rn_comparison}
\end{center}
\end{figure*}

In initial experiments, the incident count rate was entirely unconstrained, an initially surprising result that can be explained by the use of the fractional \rms normalization we used for the periodogram. This normalization rescales the periodogram such that it is independent of the mean count rate. But because the periodogram is the only summary of the data used in the model, the incident mean count rate can no longer be inferred. For the results presented in this section, we used the absolute \rms normalization \citep{uttley2001} instead, which preserves the scaling factor due to the flux in the periodogram and thus allows for accurate inference of the mean count rate as shown by the well-constrained and accurate distribution in Figure \ref{fig:rn_corner}. In line with these well-constrained posterior distribution, Figure \ref{fig:rn_post} reveals that realizations drawn from the posterior follow the observed data very closely.

We compare the SNPE-derived posterior with a posterior from classical MCMC on the periodogram. This pretends that the gaps do not exist, and is not an entirely fair comparison: both the Lomb-Scargle periodogram and the Gaussian Process-based interpolation method from \citet{wilkins2019} would almost certainly provide better performance, but come with their own limitations: accurate inference on Lomb-Scargle periodograms is not straightforward, because powers at neighbouring frequencies are not always statistically independent. The Gaussian Process interpolation can be very computationally expensive for long light curves, unless there exists a covariance function that is fast and easy to invert and that also matches the presumed underlying process well. 

We use the same model as for the SNPE, but with two re-parametrizations. Instead of the fractional \rms amplitude and the mean count rate, the model used for traditional MCMC has an amplitude for the broken power law and a parameter for the flat white noise at high frequencies (priors for all parameters in Table \ref{tab:rn_priors}). We use a standard analytical $\chi^2$ likelihood, and, where possible, the same flat priors as for the SNPE model for easier comparison.

\begin{figure*}
\begin{center}
\includegraphics[width=\textwidth]{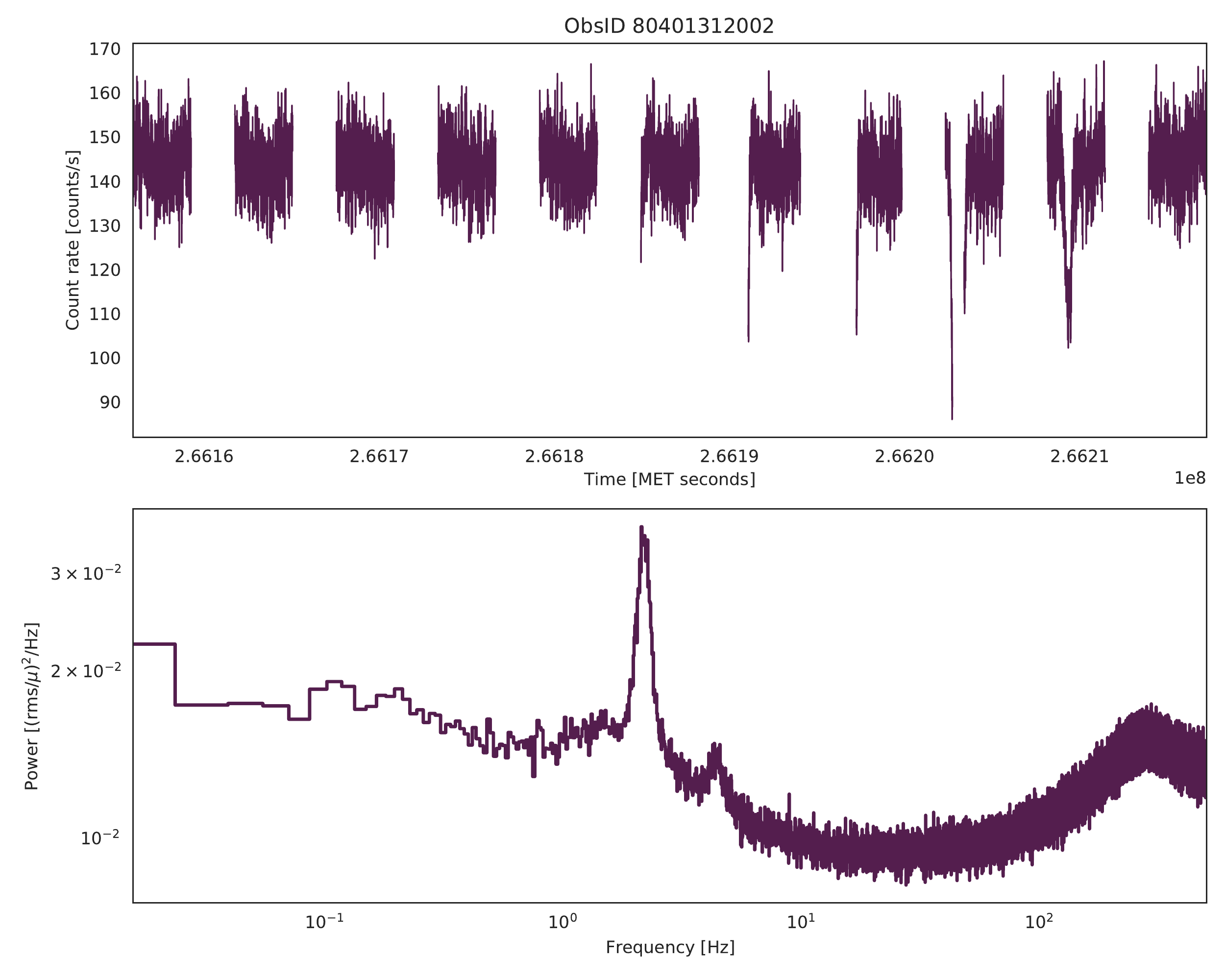}
\caption{Top panel: light curve for \nustar~observation 80401312002 of the black hole X-ray binary GRS 1915+105, binned at a 5-s resolution. Bottom panel: the averaged periodogram created from combining $530$ individual segments of $64\mathrm{s}$ duration reveals a strong QPO at $2.2\mathrm{Hz}$ and an associated harmonic at $4.4\mathrm{Hz}$. It also shows the characteristic broadband oscillatory structure typical for dead time.}
\label{fig:nustar_obs}
\end{center}
\end{figure*}

In Figure \ref{fig:rn_comparison}, we present the comparison between the simulation-based model and the analytical model without gaps for the most important parameters: the two spectral indices $\alpha_1$ and $\alpha_2$, and the break frequency $\nu_\mathrm{break}$. In both models, the low-frequency index is not well constrained, as expected given the small number of frequencies below the break frequency. Both the break frequency and the high-frequency power law index are biased away from the true value in the analytical model without gaps. Especially $\alpha_2$ is inferred to be close to a spectral index of $2.2$, and the narrow distribution excludes the true value of $3.5$. The posterior distribution inferred through SNPE, conversely, is much broader, but with a mode at the true value. The bias is less pronounced for the break frequency, but still significant. Our results suggest that SNPE can produce accurate posteriors for problems that involve data with gaps, at the cost of some precision.

\section{QPO Modelling in GRS 1915+105}\label{sec:grs1915}

To test the model on real data, we extract photon events from \nustar~observation (obsid \texttt{80401312002}) of the black hole X-ray binary GRS 1915+105. The observation has a total exposure of $26166\mathrm{s}$ and a mean count rate of $\mu_\mathrm{cr} = 143$ counts/s after summing the photons of both detector modules and taking into account GTIs. An averaged periodogram generated out of $530$ light curve segments of $64\mathrm{s}$ each reveals a strong QPO at 2.2 Hz and a harmonic at twice that frequency (Figure \ref{fig:nustar_obs}). Because we would like to explore the time-dependent structure of the QPO, we split the light curve into segments of $240\mathrm{s}$ duration, each of which is turned into a periodogram that averages 15 individual segments of $16\mathrm{s}$ duration. The periodogram is logarithmically binned with $f=0.01$ in order to reduce noise at the high frequencies. This yields a total of $207$ averaged periodograms over the full observation. Visual inspection reveals that the QPO is present over the course of the entire observation.

\begin{figure}
\begin{center}
\includegraphics[width=0.5\textwidth]{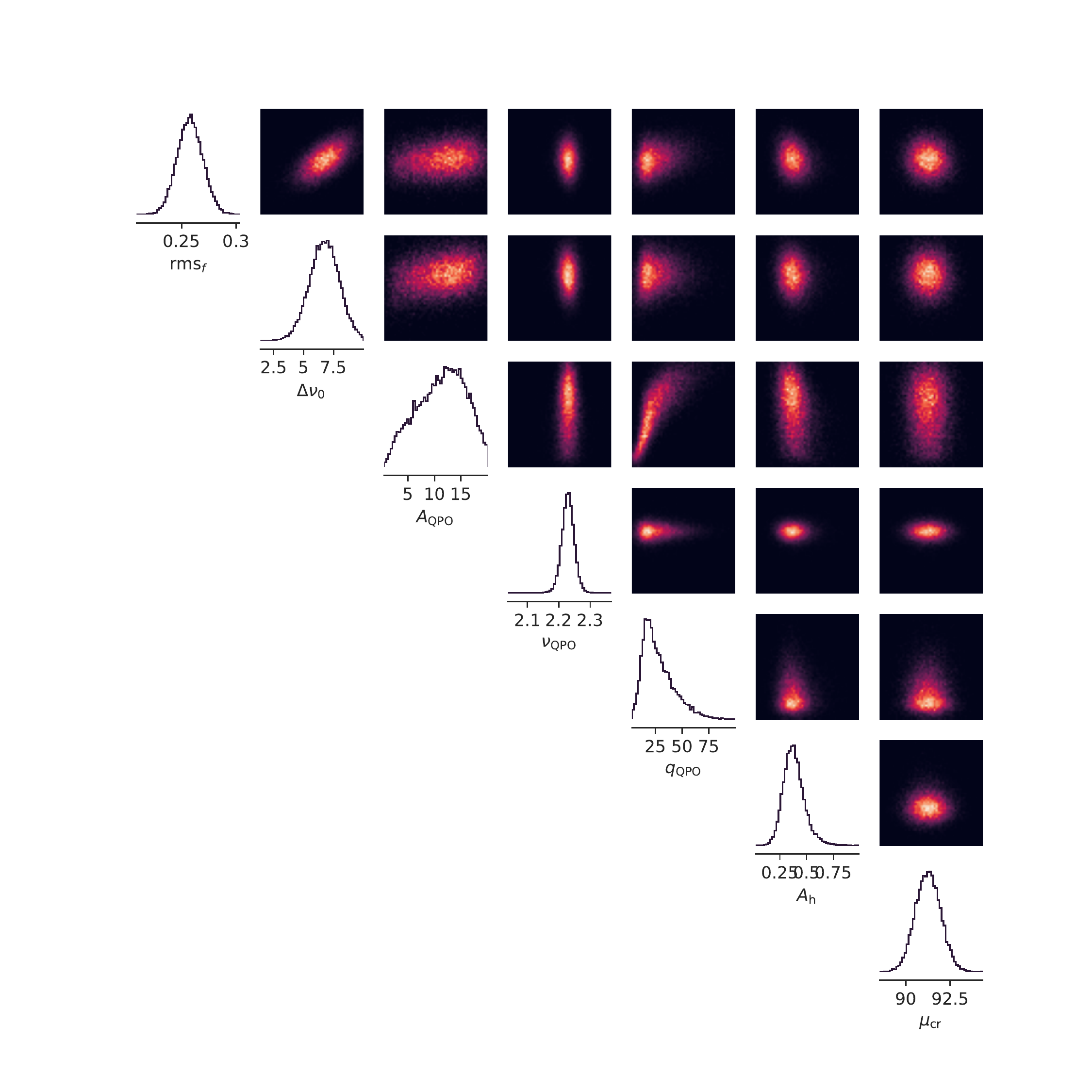}
\caption{Corner plot of the posterior probability distribution for the QPO model and the first light curve of the \nustar~observation of GRS 1915+105.}
\label{fig:grs1915_corner}
\end{center}
\end{figure}

\begin{figure}
\begin{center}
\includegraphics[width=0.5\textwidth]{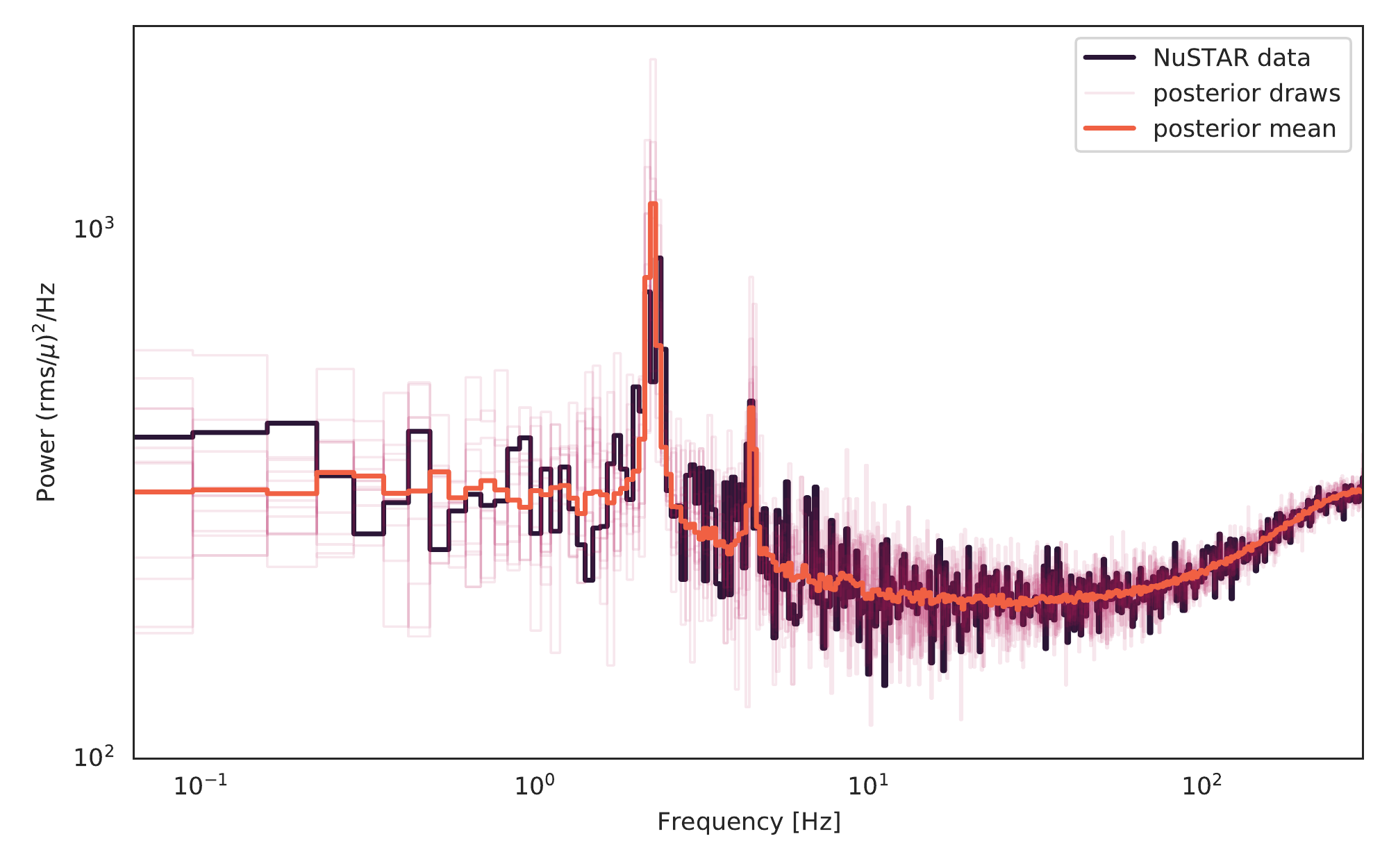}
\caption{Periodogram of the first light curve in the \nustar~observation of GRS 1915+105 (purple), along with realizations of random draws from the posterior (pink) and the posterior median (orange). The well-constrained posteriors in Figure \ref{fig:grs1915_corner} translate into realizations that generally capture all of the behaviour seen in this periodogram: the low-frequency noise, the QPO and its harmonic, and at high frequencies the imprint of dead time.}
\label{fig:grs1915_post}
\end{center}
\end{figure}

\begin{table*}[hbtp]
\renewcommand{\arraystretch}{1.3}
\footnotesize
\caption{Priors used in model for GRS 1915+105}
\begin{threeparttable} 
\begin{tabularx}{18cm}{p{2.5cm}p{7.5cm}p{4.5cm}}
\toprule
\bf{Parameter} & \bf{Meaning} & \bf{Probability Distribution} \\ \midrule
 $\mathrm{rms}_f$ & fractional \rms amplitude & $\mathrm{Uniform}(0.1,0.4)$ \\
 $\Delta\nu_0$ & FWHM of the zero-centred Lorentzian & $\mathrm{Uniform}(1,10)$ \\
 $A_\mathrm{QPO}$ & relative amplitude of the QPO & $\mathrm{Uniform}(0.5,20)$ \\
 $\nu_\mathrm{QPO}$ & QPO centroid frequency & $\mathrm{Uniform}(1.5,3.5)$ \\
 $q_\mathrm{QPO}$ & quality factor of the QPO & $\mathrm{Uniform}(3,100)$ \\
 $A_\mathrm{h}$ & relative amplitude of the harmonic & $\mathrm{Uniform}(0.1,1.0)$ \\
 $\mu_\mathrm{cr}$ & mean incident photon count rate & $\mathrm{Uniform}(500,1500)$ \\
\bottomrule
\end{tabularx}
   \begin{tablenotes}
      \item{An overview over the model parameters with their respective prior probability distributions for the model for GRS 1915+105.}
\end{tablenotes}
\end{threeparttable}
\label{tab:grs1915_priors}
\end{table*}


We build a model for the underlying power spectrum of the observed data that contains three Lorentzian components: (1) a zero-centred Lorentzian to account for the broadband noise, (2) a Lorentzian to account for the strong QPO around $2.2 \mathrm{Hz}$ and (3) a Lorentzian to account for the harmonic, with its centroid frequency required to be twice the centroid frequency of component (2) and its quality factor to be the same as component (2). We parametrize these components relative to the broadband noise component, which is fixed to an amplitude of $A_\mathrm{rn} = 1.0$, and relative to the fractional \rms amplitude of the entire spectrum. The mean incident count rate of the light curve is the final free parameter. All parameters are described with their priors in Table \ref{tab:grs1915_priors}. To simulate dead time-affected light curves, we choose the constant dead time model with $\deadtime = 0.0025\mathrm{s}$. Simulated light curves are generated with a high resolution of $dt_\mathrm{hires} = 10^{-5}\mathrm{s}$ in order to be able to accurately represent how short-term variations and dead time combine to affect the data. After removing photons due to dead time, light curves are rebinned to a resolution of $dt = 0.001\mathrm{s}$, equivalent to a Nyquist frequency of $\nu_\mathrm{Ny} = 500\mathrm{Hz}$, in order to allow the model to accurately estimate the incident count rates. 

Using this simulator, we generate $100000$ simulated periodograms from parameters drawn from the prior specified in Table \ref{tab:grs1915_priors}, which we use to train the density estimator. Based on the results in Section \ref{sec:simulations}, we choose not to include a neural network to generate summary features, but let the MAF use the periodogram directly. While a sequential estimation would be less computationally expensive, it would also require training a separate estimator for each of the $207$ individual light curves from this observation. We generate more simulations upfront and train a single MAF in order to take advantage of the amortization of this process: once trained, generating samples from the posterior is a fast and efficient process for all light curves. We also retrain the network with $10000$, $25000$, $50000$ and $75000$ simulations to understand whether we can perform similarly successful inference with fewer simulations, and find that $50000$ simulations produces posteriors of comparable shapes and widths.

In Figures \ref{fig:grs1915_corner} and \ref{fig:grs1915_post}, we show the marginalized posterior distributions and realizations drawn from the posterior probability distribution for the first of the 207 light curves. We find that the posterior density is well-constrained in a single mode for all parameters, though the quality factor has a relatively wide posterior. The comparison between the observed periodogram and posterior draws reveals some variance in the latter, expected given the noise properties of a periodogram averaged from 15 segments. The posterior median from 100 simulations traces all features in the periodogram closely and shows no inherent biases. Visual checks of the posterior samples for other light curves in the observation confirms that overall, the density estimator trained on 100,000 simulations does a good job of approximating the posteriors. 

\begin{figure*}
\begin{center}
\includegraphics[width=\textwidth]{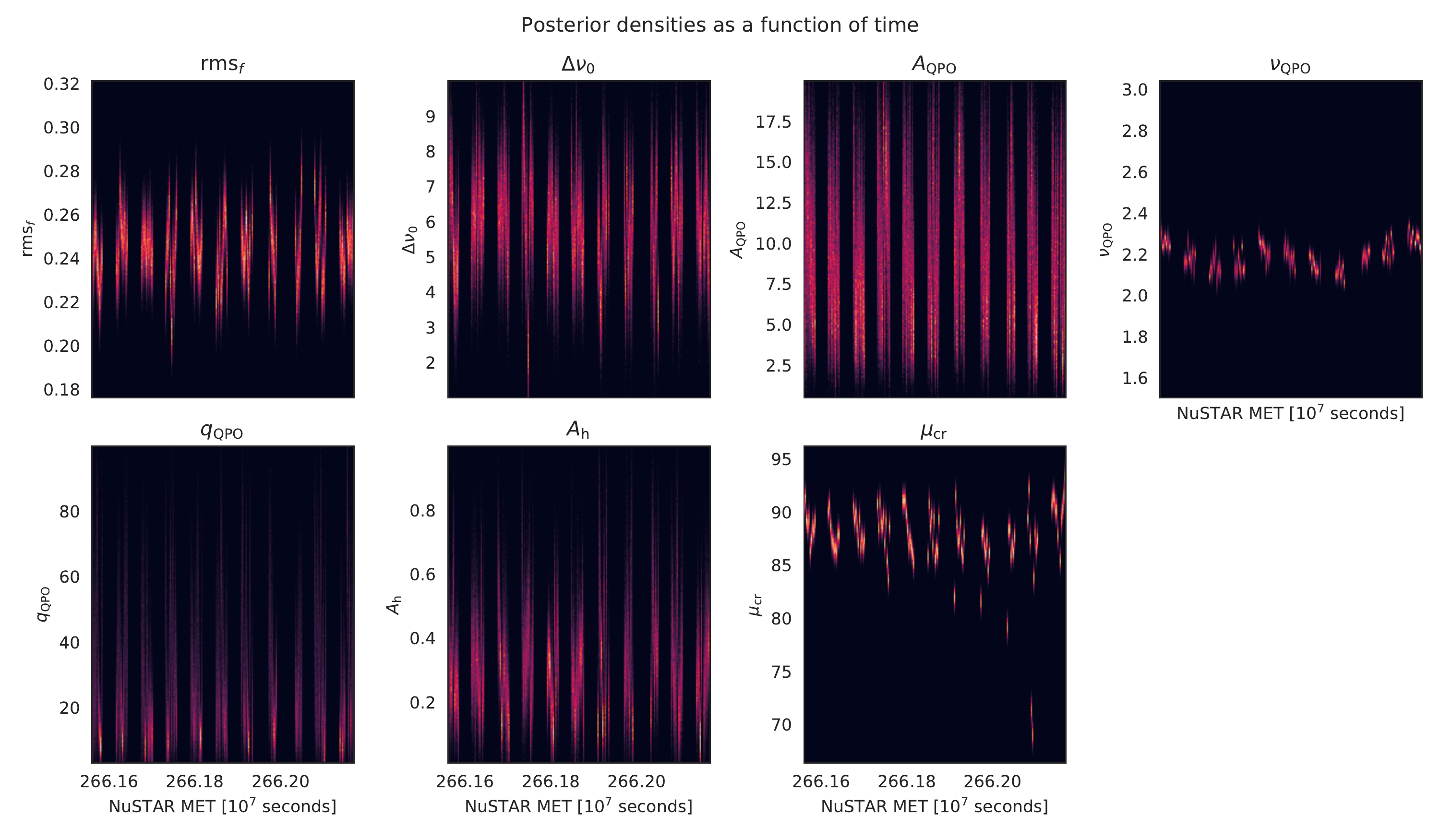}
\caption{Posterior probability densities as a function of time. For each parameter in the model, we generate histograms of the marginalized posterior probability densities out of 5000 posterior samples. For each parameter, we use the same $200$ equally spaced bins for all posteriors. The histograms are then plotted as a heatmap as a function of time (black: little or no probability, orange high probability). Black gaps are associated with gaps in the GTIs. Instrumental signatures are visible in the artificially low incident count rate estimates at the start of some GTIs.}
\label{fig:grs1915_postall}
\end{center}
\end{figure*}

In order to explore the time-dependent behaviour of the periodograms, we generate posterior samples for each of the $207$ averaged periodograms generated from this observation. Because the model is amortized, drawing posterior samples is fast, of the order of seconds for $10,000$ samples. We generate histograms with $200$ bins for the marginalized posteriors for each parameter in each segment, and then plot these histograms in a heatmap as a function of time (Figure \ref{fig:grs1915_postall}). This plot reveals significant variation in all parameters. Stark dips in the posterior for the incident count rate that occur directly at the beginning of a GTI and are likely due to instrumental effects. We also observe stochastic variability in the fractional \rms amplitude, and somewhat more structured changes in the centroid frequency of the QPO: while the other parameters appear to jitter on shorter time scales, the QPO centroid frequency seems to change on a timescale closer to the length of the total observation. However, this appearance may be related to the fact that in both our simulations and in the real data, the centroid frequency is always the parameter with the narrowest posterior density, thus we cannot say whether there is an intrinsic difference between variability in different parameters. To the eye, this variation may appear near-periodic, though we caution the reader that this might be an illusion generated by an otherwise stochastic process \citep[e.g.][]{press1978}. 

\begin{figure*}
\begin{center}
\includegraphics[width=\textwidth]{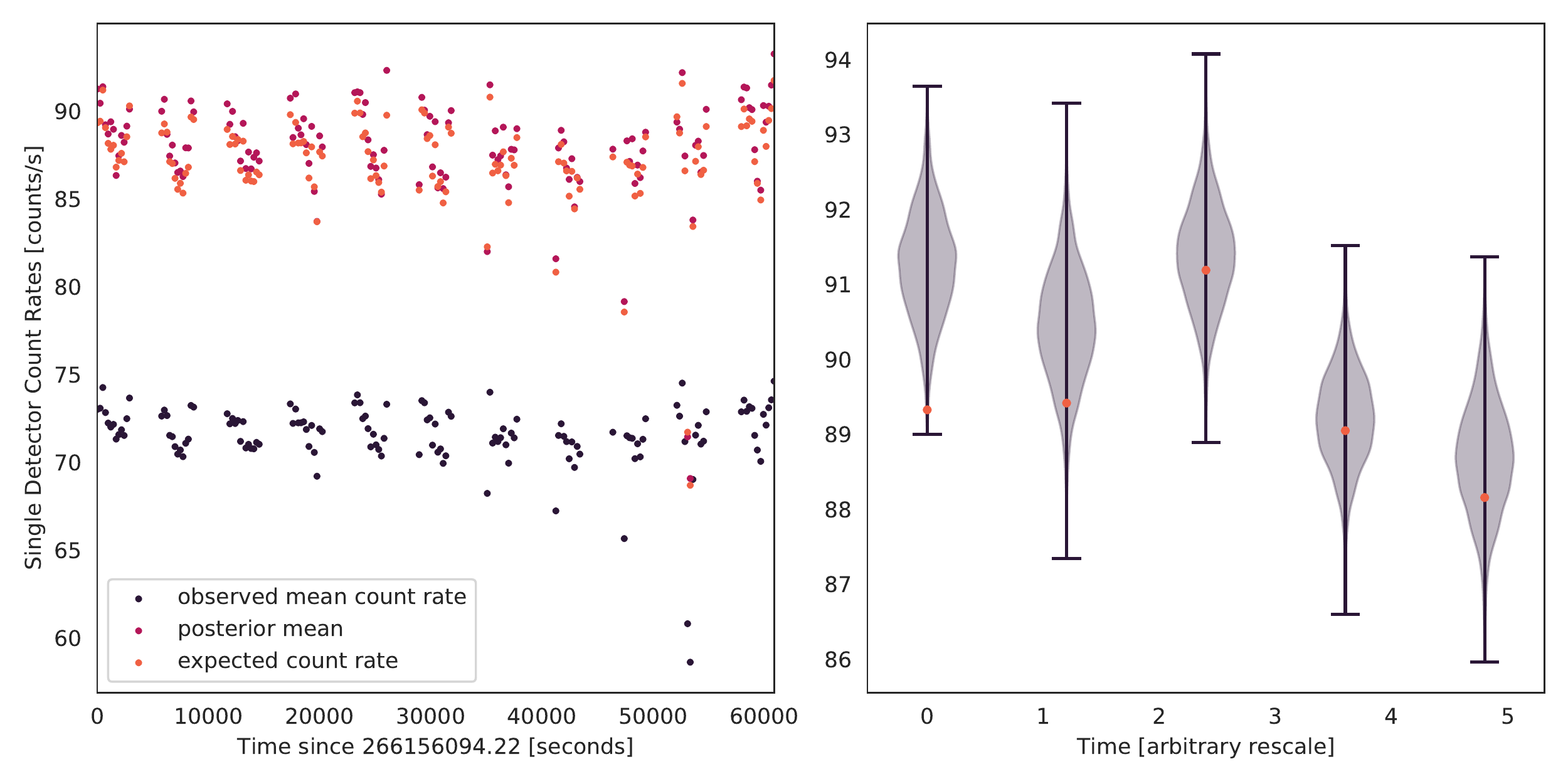}
\caption{Left: Comparison of the observed mean count rate (purple), the mean of the posterior probability distribution (pink) and the expected mean count rate calculated using the observed count rate and \nustar~constant dead time using Equation \ref{eqn:exp_deadtime} for each of the 207 light curves in \nustar~observation \texttt{80401312002}. Right: violin plot of the posterior probability densities for the incident mean count rate (purple), and the expected mean count rate (orange) as in the left-hand plot, for the first five light curves.}
\label{fig:grs1915_crcomp}
\end{center}
\end{figure*}

As a check on the validity of the model, we compare the inferred incident count rates with the observed mean count rate in each light curve, and with a commonly-used analytical estimate for the incident count rate, given by

\begin{equation} \label{eqn:exp_deadtime}
    \mu_\mathrm{exp} = \frac{\mu_\mathrm{obs}}{1 - \deadtime\mu_\mathrm{obs}} \; ,
\end{equation}

\noindent where $\mu_\mathrm{obs}$ is the observed mean count rate and $\deadtime$ is the dead time. Figure \ref{fig:grs1915_crcomp} presents the comparison between observed count rates, expected count rates, and the mean of the posterior pdf for each of the light curves as a function of time. Overall, both the analytic estimate and the posterior means are significantly higher than the observed count rates, as expected for a bright light curve with a significant loss of photons. While the posterior means and the analytic estimate appear very similar, the posterior means are systematically higher by $\sim 0.75 \mathrm{counts}/\mathrm{s}$. This may indicate that the analytic estimate is a slight underestimation. Both analytic estimate and posterior means used a constant dead time model; if variable dead time were consistently taken into account, we would expect the incident count rate somewhat higher, as was the case in Section \ref{sec:variable}. Because posterior means are only an incomplete representation of a full posterior, we also present violin plots of the posterior for the first five light curves along with the analytical estimate as a comparison. While overall, the distributions and the analytical estimates agree, the latter seems to generally fall into the lower half of the distribution, in line with the overall observation that the posterior favours slightly higher incident mean count rates. 

\section{Discussion}\label{sec:discussion}

Problems where an analytical likelihood is inaccessible or where the data is affected by instrumental effects that are hard to incorporate in a parametric model are common in science. X-ray astronomy is no exception, and dead time has been a long-standing issue for timing studies in this field. Our work here presents a powerful new approach to mitigating instrumental effects like dead time, based on recent advances in density estimation via neural networks, which have dramatically improved the efficiency and accuracy of simulation-based inference. Simulations to assess instrumental effects are not new in astronomy. However, simulation-based inference as presented here embeds these simulations into the framework of probabilistic modelling and allows for a direct estimation of posterior probability distributions from these simulations.

We used dead time in \nustar~as an example, but in some ways, \nustar~is the least interesting instrument for this kind of approach. Its two identical detectors make other, faster methods like the FAD accessible, though we have also shown that the dead time distribution in \nustar~does matter compared to the constant value often assumed. In studies that require extremely accurate measurements of the incident flux, the model proposed here, especially when drawing individual dead time intervals from the observed distribution, is likely to produce less biased results, though this will also depend on the strength of the variability in question (biases are most likely strongest for very high fractional rms amplitudes; \citealt{bachetti2018}). However, there are instruments for which alternative approaches are not available, because the instrument comprises a single detector, or because the dead time process itself is more complex (e.g.~energy-dependent dead time). In these cases, simulations might be the only option to assess and mitigate dead time effects. 

Traditionally, simulation-based inference required the careful construction of handcrafted summary features from the data to enable the algorithm to efficiently construct posteriors. For a relatively simple problem with few parameters, we have shown that SNPE works impressively well on the periodograms themselves. We note, however, that the periodogram in fact \textit{is} a summary of the data, albeit a noisy and incomplete one. For stochastic processes common in accreting sources like GRS 1915+105, it is a better estimator than the raw light curves would be, though our experience with the example of a red noise process with gaps and without dead time also shows that a careful construction of the summaries used is still important. Because the mean flux in a light curve imposes a constant scaling factor on the periodogram that is divided out in the fractional \rms normalization, a model that uses this normalization to construct the periodogram as a summary will not be able to infer the incident count rate in the light curve, because there is no information available in the summary to do. Interestingly, we only noticed this in the model \textit{without} dead time, despite the fact that all previous simulations in Section \ref{sec:simulations} also employ the same normalization. We conclude that in these other cases, the incident mean count rate was actually inferred through the dead time process itself. Because the effect of dead time is flux-dependent, the model can use the shape of the periodogram across a wide range of frequencies to estimate the incident flux, despite no other information about the flux being available.

In this paper, we focused on introducing and validating simulation-based inference using relatively simple toy problems to showcase the approach, but the method admits much more complex problems and data sets. In principle, one might combine different, more comprehensive summaries of the data, such as time lags, covariance-energy spectra, phase-resolved spectroscopy, or \rms-flux spectra in order to perform joint probabilistic inference given an underlying physical model for the effects observed in the data. Because new, neural-network-based methods for simulation-based inference scale well to higher dimensions, and because they also allow for the use of intrinsically stochastic models, limitations to this approach come primarily from the ability to simulate the process in question. 

While in the past, simulation-based inference was extremely computationally expensive, the sequential approach proposed by \citet{greenberg2019} makes it possible to construct accurate posteriors with a few thousand simulations; less than or at most equal to what many MCMC algorithms commonly used in astronomy require. Amortized inference without the sequential components is significantly more computationally expensive, but depending on the problem at hand, that upfront expense may be justified if it enables inference across many different data sets later on, as we have shown on the GRS 1915+105 data. While we do not find that summaries constructed through convolutional neural networks improve our results, and posterior inference seems to work well on the periodograms from GRS 1915+105 data without such summaries, it is possible that other sources may benefit from feature extraction with convolutional neural networks. In general, while our results provide a proof-of-concept that SBI with SNPE enables reliable inference in the presence of dead time, our modelling choices should not be seen as prescriptive, especially when applying this approach to sources with significantly different light curves and periodograms, which may require additional experimentation and hyperparameter optimization.

We caution that a simulator based on a physical model might be computationally expensive. For the problem considered here, we found that simulating whole light curves of more than a few thousand seconds becomes very slow because of the simultaneous requirement of a time resolution smaller than the typical time scale of the dead time process. This leads to an excessively large number of frequencies to be inverse Fourier-transformed into a light curve, and consequently to a slow simulator. However, simulations can be pre-recorded, and easily parallelized. Amortized inference on multiple observations only works if the observations all have the same overall properties as the simulated data: that is, they must have the same length and time resolution in order to be comparable to the simulated data. Finally, while neural networks have been shown to be extremely efficient and flexible estimators of densities, they are also opaque models that are difficult to interpret. Understanding when they fail, or how well they estimate the target posterior density in practice can be difficult. We performed extensive simulations to understand the model's performance on a relatively simple problem with dead time. We also employed posterior predictive checking through comparisons between the observed periodogram and realizations of the posteriors as well as the posterior median to identify possible biases in the model.  

The fast posterior sampling using the amortized model has additional advantages, because it opens up spectral timing to hierarchical inference. \citet{rodrigues2021} recently introduced an algorithm for SNPE in a hierarchical modelling context. In the context of the GRS 1915+105 data presented above, a hierarchical model could help constrain the variability properties of the QPO centroid frequency. Given a (stochastic or periodic) function for the evolution of the QPO centroid frequency in time, a hierarchical model might define the parameters of that function as population-level parameters, and recast the centroid frequency for each individual segment as latent variables. In a hierarchical framework, these parameters are jointly inferred. This provides a powerful tool for future spectral timing studies of accreting sources.

\section*{Acknowledgements}

D.H. is supported by the Women In Science Excel (WISE) programme of the Netherlands Organisation for Scientific Research (NWO). DH acknowledges partial support from NASA grant 80NSSC18K0425. DH acknowledges support from the DiRAC Institute in the Department of Astronomy at the University of Washington. The DiRAC Institute is supported through generous gifts from the Charles and Lisa Simonyi Fund for Arts and Sciences, and the Washington Research Foundation.
M.B. is supported in part by PRIN INAF 2019 -- \textit{Spectempolar! Timing analysis in the era of high-throughput photon detectors.} 
The authors are grateful to Adam Ingram for providing the \nustar~observation used in this manuscript. D.H. thanks Kyle Cranmer and Gilles Louppe for a number of early conversations that were instrumental to the ideation of this project.
The authors acknowledge the Deep Skies Lab as a community of multi-domain experts and collaborators who have facilitated an environment of open discussion, idea-generation, and collaboration. This community was important for the development of this project.

\section*{Data Availability}

 The \nustar~observation used in this manuscript is publicly available from HEASARC\footnote{\url{https://heasarc.gsfc.nasa.gov/}}. Pre-recorded simulation runs are available for download on  \href{http://doi.org/10.5281/zenodo.4670161}{Zenodo}, and all code related to the project is publicly accessible on its  \href{https://github.com/dhuppenkothen/DeadTimeSBI}{GitHub repository}.



\bibliographystyle{mnras}
\bibliography{deadtime_sbi} 

\begin{thebibliography}{}
\makeatletter
\relax
\def\mn@urlcharsother{\let\do\@makeother \do\$\do\&\do\#\do\^\do\_\do\%\do\~}
\def\mn@doi{\begingroup\mn@urlcharsother \@ifnextchar [ {\mn@doi@}
  {\mn@doi@[]}}
\def\mn@doi@[#1]#2{\def\@tempa{#1}\ifx\@tempa\@empty \href
  {http://dx.doi.org/#2} {doi:#2}\else \href {http://dx.doi.org/#2} {#1}\fi
  \endgroup}
\def\mn@eprint#1#2{\mn@eprint@#1:#2::\@nil}
\def\mn@eprint@arXiv#1{\href {http://arxiv.org/abs/#1} {{\tt arXiv:#1}}}
\def\mn@eprint@dblp#1{\href {http://dblp.uni-trier.de/rec/bibtex/#1.xml}
  {dblp:#1}}
\def\mn@eprint@#1:#2:#3:#4\@nil{\def\@tempa {#1}\def\@tempb {#2}\def\@tempc
  {#3}\ifx \@tempc \@empty \let \@tempc \@tempb \let \@tempb \@tempa \fi \ifx
  \@tempb \@empty \def\@tempb {arXiv}\fi \@ifundefined
  {mn@eprint@\@tempb}{\@tempb:\@tempc}{\expandafter \expandafter \csname
  mn@eprint@\@tempb\endcsname \expandafter{\@tempc}}}

\bibitem[\protect\citeauthoryear{{Alsing}, {Charnock}, {Feeney}  \&
  {Wandelt}}{{Alsing} et~al.}{2019}]{2019MNRAS.488.4440A}
{Alsing} J.,  {Charnock} T.,  {Feeney} S.,   {Wandelt} B.,  2019, \mn@doi
  [\mnras] {10.1093/mnras/stz1960}, \href
  {https://ui.adsabs.harvard.edu/abs/2019MNRAS.488.4440A} {488, 4440}

\bibitem[\protect\citeauthoryear{{Aufort}, {Ciesla}, {Pudlo}  \&
  {Buat}}{{Aufort} et~al.}{2020}]{2020A&A...635A.136A}
{Aufort} G.,  {Ciesla} L.,  {Pudlo} P.,   {Buat} V.,  2020, \mn@doi [\aap]
  {10.1051/0004-6361/201936788}, \href
  {https://ui.adsabs.harvard.edu/abs/2020A&A...635A.136A} {635, A136}

\bibitem[\protect\citeauthoryear{Bachetti \& Huppenkothen}{Bachetti \&
  Huppenkothen}{2018}]{bachetti2018}
Bachetti M.,  Huppenkothen D.,  2018, \mn@doi [ApJL]
  {10.3847/2041-8213/aaa83b}, 853, L21

\bibitem[\protect\citeauthoryear{Bachetti et~al.,}{Bachetti
  et~al.}{2015}]{Bachetti+15}
Bachetti M.,  et~al., 2015, ApJ, 800, 109

\bibitem[\protect\citeauthoryear{{Barret} et~al.,}{{Barret}
  et~al.}{2018}]{xifuspie}
{Barret} D.,  et~al., 2018, in {den Herder} J.-W.~A.,  {Nikzad} S.,
  {Nakazawa} K.,  eds,  Society of Photo-Optical Instrumentation Engineers
  (SPIE) Conference Series Vol. 10699, Space Telescopes and Instrumentation
  2018: Ultraviolet to Gamma Ray. p. 106991G (\mn@eprint {arXiv} {1807.06092}),
  \mn@doi{10.1117/12.2312409}

\bibitem[\protect\citeauthoryear{Beaumont, Zhang  \& Balding}{Beaumont
  et~al.}{2002}]{beaumont2002approximate}
Beaumont M.~A.,  Zhang W.,   Balding D.~J.,  2002, Genetics, 162, 2025

\bibitem[\protect\citeauthoryear{{Bryson}, {Coughlin}, {Kunimoto}  \&
  {Mullally}}{{Bryson} et~al.}{2020}]{2020AJ....160..200B}
{Bryson} S.,  {Coughlin} J.~L.,  {Kunimoto} M.,   {Mullally} S.~E.,  2020,
  \mn@doi [\aj] {10.3847/1538-3881/abb316}, \href
  {https://ui.adsabs.harvard.edu/abs/2020AJ....160..200B} {160, 200}

\bibitem[\protect\citeauthoryear{{Cheng}, {Price-Jones}  \& {Bovy}}{{Cheng}
  et~al.}{2020}]{2020arXiv201009721C}
{Cheng} C.~M.,  {Price-Jones} N.,   {Bovy} J.,  2020, arXiv e-prints, \href
  {https://ui.adsabs.harvard.edu/abs/2020arXiv201009721C} {p. arXiv:2010.09721}

\bibitem[\protect\citeauthoryear{{Cisewski-Kehe}, {Weller}  \&
  {Schafer}}{{Cisewski-Kehe} et~al.}{2019}]{2019arXiv190411306C}
{Cisewski-Kehe} J.,  {Weller} G.,   {Schafer} C.,  2019, arXiv e-prints, \href
  {https://ui.adsabs.harvard.edu/abs/2019arXiv190411306C} {p. arXiv:1904.11306}

\bibitem[\protect\citeauthoryear{Cranmer, Brehmer  \& Louppe}{Cranmer
  et~al.}{2020}]{cranmer2020}
Cranmer K.,  Brehmer J.,   Louppe G.,  2020, \mn@doi [Proceedings of the
  National Academy of Sciences] {10.1073/pnas.1912789117}, 117, 30055

\bibitem[\protect\citeauthoryear{{Cutri} et~al.,}{{Cutri}
  et~al.}{2003}]{2003yCat.2246....0C}
{Cutri} R.~M.,  et~al., 2003, VizieR Online Data Catalog, \href
  {https://ui.adsabs.harvard.edu/abs/2003yCat.2246....0C} {p. II/246}

\bibitem[\protect\citeauthoryear{Diggle \& Gratton}{Diggle \&
  Gratton}{1984}]{diggle1984monte}
Diggle P.~J.,  Gratton R.~J.,  1984, Journal of the Royal Statistical Society:
  Series B (Methodological), 46, 193

\bibitem[\protect\citeauthoryear{Durkan, Murray  \& Papamakarios}{Durkan
  et~al.}{2020}]{durkan2020contrastive}
Durkan C.,  Murray I.,   Papamakarios G.,  2020, in International Conference on
  Machine Learning. pp 2771--2781

\bibitem[\protect\citeauthoryear{{Enzi}, {Vegetti}, {Despali}, {Hsueh}  \&
  {Metcalf}}{{Enzi} et~al.}{2020}]{2020MNRAS.496.1718E}
{Enzi} W.,  {Vegetti} S.,  {Despali} G.,  {Hsueh} J.-W.,   {Metcalf} R.~B.,
  2020, \mn@doi [\mnras] {10.1093/mnras/staa1224}, \href
  {https://ui.adsabs.harvard.edu/abs/2020MNRAS.496.1718E} {496, 1718}

\bibitem[\protect\citeauthoryear{Fender, Belloni  \& Gallo}{Fender
  et~al.}{2004}]{fender2004}
Fender R.~P.,  Belloni T.~M.,   Gallo E.,  2004, \mn@doi [MNRAS]
  {10.1111/j.1365-2966.2004.08384.x}, 355, 1105

\bibitem[\protect\citeauthoryear{{Foreman-Mackey}, {Hogg}, {Lang}  \&
  {Goodman}}{{Foreman-Mackey} et~al.}{2013}]{emcee}
{Foreman-Mackey} D.,  {Hogg} D.~W.,  {Lang} D.,   {Goodman} J.,  2013, \mn@doi
  [\pasp] {10.1086/670067}, \href
  {https://ui.adsabs.harvard.edu/abs/2013PASP..125..306F} {125, 306}

\bibitem[\protect\citeauthoryear{Greenberg, Nonnenmacher  \& Macke}{Greenberg
  et~al.}{2019}]{greenberg2019}
Greenberg D.,  Nonnenmacher M.,   Macke J.,  2019, in Chaudhuri K.,
  Salakhutdinov R.,  eds,  Proceedings of Machine Learning Research Vol. 97,
  Proceedings of the 36th International Conference on Machine Learning. PMLR,
  pp 2404--2414, \url {http://proceedings.mlr.press/v97/greenberg19a.html}

\bibitem[\protect\citeauthoryear{{Hahn}, {Vakili}, {Walsh}, {Hearin}, {Hogg}
  \& {Campbell}}{{Hahn} et~al.}{2017}]{2017MNRAS.469.2791H}
{Hahn} C.,  {Vakili} M.,  {Walsh} K.,  {Hearin} A.~P.,  {Hogg} D.~W.,
  {Campbell} D.,  2017, \mn@doi [\mnras] {10.1093/mnras/stx894}, \href
  {https://ui.adsabs.harvard.edu/abs/2017MNRAS.469.2791H} {469, 2791}

\bibitem[\protect\citeauthoryear{{Harrison} et~al.,}{{Harrison}
  et~al.}{2013}]{harrison2013}
{Harrison} F.~A.,  et~al., 2013, \mn@doi [\apj] {10.1088/0004-637X/770/2/103},
  \href {https://ui.adsabs.harvard.edu/abs/2013ApJ...770..103H} {770, 103}

\bibitem[\protect\citeauthoryear{Hastings}{Hastings}{1970}]{hastings1970}
Hastings W.~K.,  1970, \mn@doi [Biometrika] {10.1093/biomet/57.1.97}, 57, 97

\bibitem[\protect\citeauthoryear{{He} et~al.,}{{He}
  et~al.}{2020}]{2020arXiv201013221H}
{He} Q.,  et~al., 2020, arXiv e-prints, \href
  {https://ui.adsabs.harvard.edu/abs/2020arXiv201013221H} {p. arXiv:2010.13221}

\bibitem[\protect\citeauthoryear{{Herbel}, {Kacprzak}, {Amara}, {Refregier},
  {Bruderer}  \& {Nicola}}{{Herbel} et~al.}{2017}]{2017JCAP...08..035H}
{Herbel} J.,  {Kacprzak} T.,  {Amara} A.,  {Refregier} A.,  {Bruderer} C.,
  {Nicola} A.,  2017, \mn@doi [\jcap] {10.1088/1475-7516/2017/08/035}, \href
  {https://ui.adsabs.harvard.edu/abs/2017JCAP...08..035H} {2017, 035}

\bibitem[\protect\citeauthoryear{Hermans, Begy  \& Louppe}{Hermans
  et~al.}{2019}]{hermans2019likelihood}
Hermans J.,  Begy V.,   Louppe G.,  2019, arXiv preprint arXiv:1903.04057, 10

\bibitem[\protect\citeauthoryear{{Hermans}, {Banik}, {Weniger}, {Bertone}  \&
  {Louppe}}{{Hermans} et~al.}{2020}]{2020arXiv201114923H}
{Hermans} J.,  {Banik} N.,  {Weniger} C.,  {Bertone} G.,   {Louppe} G.,  2020,
  arXiv e-prints, \href {https://ui.adsabs.harvard.edu/abs/2020arXiv201114923H}
  {p. arXiv:2011.14923}

\bibitem[\protect\citeauthoryear{{Hsu}, {Ford}  \& {Terrien}}{{Hsu}
  et~al.}{2020}]{2020MNRAS.498.2249H}
{Hsu} D.~C.,  {Ford} E.~B.,   {Terrien} R.,  2020, \mn@doi [\mnras]
  {10.1093/mnras/staa2391}, \href
  {https://ui.adsabs.harvard.edu/abs/2020MNRAS.498.2249H} {498, 2249}

\bibitem[\protect\citeauthoryear{{Huppenkothen} \& {Bachetti}}{{Huppenkothen}
  \& {Bachetti}}{2018}]{huppenkothen2018}
{Huppenkothen} D.,  {Bachetti} M.,  2018, \mn@doi [\apjs]
  {10.3847/1538-4365/aabe38}, \href
  {https://ui.adsabs.harvard.edu/abs/2018ApJS..236...13H} {236, 13}

\bibitem[\protect\citeauthoryear{{Huppenkothen} et~al.,}{{Huppenkothen}
  et~al.}{2019}]{stingraypaper}
{Huppenkothen} D.,  et~al., 2019, \mn@doi [\apj] {10.3847/1538-4357/ab258d},
  \href {https://ui.adsabs.harvard.edu/abs/2019ApJ...881...39H} {881, 39}

\bibitem[\protect\citeauthoryear{{Ingram}}{{Ingram}}{2019}]{ingram2019}
{Ingram} A.,  2019, \mn@doi [\mnras] {10.1093/mnras/stz2409}, \href
  {https://ui.adsabs.harvard.edu/abs/2019MNRAS.489.3927I} {489, 3927}

\bibitem[\protect\citeauthoryear{{Izbicki}, {Lee}  \& {Pospisil}}{{Izbicki}
  et~al.}{2018}]{npe4}
{Izbicki} R.,  {Lee} A.~B.,   {Pospisil} T.,  2018, arXiv e-prints, \href
  {https://ui.adsabs.harvard.edu/abs/2018arXiv180505480I} {p. arXiv:1805.05480}

\bibitem[\protect\citeauthoryear{{Jennings} \& {Madigan}}{{Jennings} \&
  {Madigan}}{2017}]{jennings2017}
{Jennings} E.,  {Madigan} M.,  2017, \mn@doi [Astronomy and Computing]
  {10.1016/j.ascom.2017.01.001}, \href
  {https://ui.adsabs.harvard.edu/abs/2017A&C....19...16J} {19, 16}

\bibitem[\protect\citeauthoryear{{Jennings}, {Wolf}  \& {Sako}}{{Jennings}
  et~al.}{2016}]{jennings2016}
{Jennings} E.,  {Wolf} R.,   {Sako} M.,  2016, arXiv e-prints, \href
  {https://ui.adsabs.harvard.edu/abs/2016arXiv161103087J} {p. arXiv:1611.03087}

\bibitem[\protect\citeauthoryear{{Jimenez Rezende} \& {Mohamed}}{{Jimenez
  Rezende} \& {Mohamed}}{2015}]{npe1}
{Jimenez Rezende} D.,  {Mohamed} S.,  2015, arXiv e-prints, \href
  {https://ui.adsabs.harvard.edu/abs/2015arXiv150505770J} {p. arXiv:1505.05770}

\bibitem[\protect\citeauthoryear{{Kacprzak} et~al.,}{{Kacprzak}
  et~al.}{2020}]{2020PhRvD.101h2003K}
{Kacprzak} T.,  et~al., 2020, \mn@doi [\prd] {10.1103/PhysRevD.101.082003},
  \href {https://ui.adsabs.harvard.edu/abs/2020PhRvD.101h2003K} {101, 082003}

\bibitem[\protect\citeauthoryear{{Kunimoto} \& {Bryson}}{{Kunimoto} \&
  {Bryson}}{2021}]{2021AJ....161...69K}
{Kunimoto} M.,  {Bryson} S.,  2021, \mn@doi [\aj] {10.3847/1538-3881/abd2c1},
  \href {https://ui.adsabs.harvard.edu/abs/2021AJ....161...69K} {161, 69}

\bibitem[\protect\citeauthoryear{{Kunimoto} \& {Matthews}}{{Kunimoto} \&
  {Matthews}}{2020}]{2020AJ....159..248K}
{Kunimoto} M.,  {Matthews} J.~M.,  2020, \mn@doi [\aj]
  {10.3847/1538-3881/ab88b0}, \href
  {https://ui.adsabs.harvard.edu/abs/2020AJ....159..248K} {159, 248}

\bibitem[\protect\citeauthoryear{{Leclercq}, {Enzi}, {Jasche}  \&
  {Heavens}}{{Leclercq} et~al.}{2019}]{2019MNRAS.490.4237L}
{Leclercq} F.,  {Enzi} W.,  {Jasche} J.,   {Heavens} A.,  2019, \mn@doi
  [\mnras] {10.1093/mnras/stz2718}, \href
  {https://ui.adsabs.harvard.edu/abs/2019MNRAS.490.4237L} {490, 4237}

\bibitem[\protect\citeauthoryear{{List} \& {Lewis}}{{List} \&
  {Lewis}}{2020}]{2020MNRAS.493.5913L}
{List} F.,  {Lewis} G.~F.,  2020, \mn@doi [\mnras] {10.1093/mnras/staa523},
  \href {https://ui.adsabs.harvard.edu/abs/2020MNRAS.493.5913L} {493, 5913}

\bibitem[\protect\citeauthoryear{{Lomb}}{{Lomb}}{1976}]{lomb1976}
{Lomb} N.~R.,  1976, \mn@doi [Astrophysics and Space Sciences]
  {10.1007/BF00648343}, \href
  {https://ui.adsabs.harvard.edu/abs/1976Ap&SS..39..447L} {39, 447}

\bibitem[\protect\citeauthoryear{{Lueckmann}, {Goncalves}, {Bassetto},
  {{\"O}cal}, {Nonnenmacher}  \& {Macke}}{{Lueckmann}
  et~al.}{2017}]{2017arXiv171101861L}
{Lueckmann} J.-M.,  {Goncalves} P.~J.,  {Bassetto} G.,  {{\"O}cal} K.,
  {Nonnenmacher} M.,   {Macke} J.~H.,  2017, arXiv e-prints, \href
  {https://ui.adsabs.harvard.edu/abs/2017arXiv171101861L} {p. arXiv:1711.01861}

\bibitem[\protect\citeauthoryear{Lueckmann, Bassetto, Karaletsos  \&
  Macke}{Lueckmann et~al.}{2019}]{lueckmann2019likelihood}
Lueckmann J.-M.,  Bassetto G.,  Karaletsos T.,   Macke J.~H.,  2019, in
  Symposium on Advances in Approximate Bayesian Inference. pp 32--53

\bibitem[\protect\citeauthoryear{{Lueckmann}, {Boelts}, {Greenberg},
  {Gon{\c{c}}alves}  \& {Macke}}{{Lueckmann}
  et~al.}{2021}]{2021arXiv210104653L}
{Lueckmann} J.-M.,  {Boelts} J.,  {Greenberg} D.~S.,  {Gon{\c{c}}alves} P.~J.,
   {Macke} J.~H.,  2021, arXiv e-prints, \href
  {https://ui.adsabs.harvard.edu/abs/2021arXiv210104653L} {p. arXiv:2101.04653}

\bibitem[\protect\citeauthoryear{{Meegan} et~al.,}{{Meegan}
  et~al.}{2009}]{meegan2009}
{Meegan} C.,  et~al., 2009, \mn@doi [\apj] {10.1088/0004-637X/702/1/791}, \href
  {https://ui.adsabs.harvard.edu/abs/2009ApJ...702..791M} {702, 791}

\bibitem[\protect\citeauthoryear{{Metropolis}, {Rosenbluth}, {Rosenbluth},
  {Teller}  \& {Teller}}{{Metropolis} et~al.}{1953}]{metropolis1953}
{Metropolis} N.,  {Rosenbluth} A.~W.,  {Rosenbluth} M.~N.,  {Teller} A.~H.,
  {Teller} E.,  1953, \mn@doi [\jcp] {10.1063/1.1699114}, \href
  {https://ui.adsabs.harvard.edu/abs/1953JChPh..21.1087M} {21, 1087}

\bibitem[\protect\citeauthoryear{{Mor}, {Robin}, {Figueras}, {Roca-F{\`a}brega}
   \& {Luri}}{{Mor} et~al.}{2019}]{2019A&A...624L...1M}
{Mor} R.,  {Robin} A.~C.,  {Figueras} F.,  {Roca-F{\`a}brega} S.,   {Luri} X.,
  2019, \mn@doi [\aap] {10.1051/0004-6361/201935105}, \href
  {https://ui.adsabs.harvard.edu/abs/2019A&A...624L...1M} {624, L1}

\bibitem[\protect\citeauthoryear{{Morris}}{{Morris}}{2020}]{2020ApJ...893...67M}
{Morris} B.~M.,  2020, \mn@doi [\apj] {10.3847/1538-4357/ab79a0}, \href
  {https://ui.adsabs.harvard.edu/abs/2020ApJ...893...67M} {893, 67}

\bibitem[\protect\citeauthoryear{{Paige} \& {Wood}}{{Paige} \&
  {Wood}}{2016}]{npe2}
{Paige} B.,  {Wood} F.,  2016, arXiv e-prints, \href
  {https://ui.adsabs.harvard.edu/abs/2016arXiv160206701P} {p. arXiv:1602.06701}

\bibitem[\protect\citeauthoryear{Papamakarios \& Murray}{Papamakarios \&
  Murray}{2016}]{papamakarios2016fast}
Papamakarios G.,  Murray I.,  2016, arXiv preprint arXiv:1605.06376

\bibitem[\protect\citeauthoryear{Papamakarios, Pavlakou  \&
  Murray}{Papamakarios et~al.}{2017}]{papamakarios2017}
Papamakarios G.,  Pavlakou T.,   Murray I.,  2017, in Proceedings of the 31st
  International Conference on Neural Information Processing Systems. NIPS'17.
Curran Associates Inc., Red Hook, NY, USA, p. 2335–2344

\bibitem[\protect\citeauthoryear{Papamakarios, Sterratt  \&
  Murray}{Papamakarios et~al.}{2019}]{papamakarios2019sequential}
Papamakarios G.,  Sterratt D.,   Murray I.,  2019, in The 22nd International
  Conference on Artificial Intelligence and Statistics. pp 837--848

\bibitem[\protect\citeauthoryear{{Peille}}{{Peille}}{2016}]{peillethesis}
{Peille} P.,  2016, PhD thesis, Universit\'e Toulouse 3 Paul Sabatier (UT3 Paul
  Sabatier)

\bibitem[\protect\citeauthoryear{{Press}}{{Press}}{1978}]{press1978}
{Press} W.~H.,  1978, Comments on Astrophysics, \href
  {https://ui.adsabs.harvard.edu/abs/1978ComAp...7..103P} {7, 103}

\bibitem[\protect\citeauthoryear{Pritchard, Seielstad, Perez-Lezaun  \&
  Feldman}{Pritchard et~al.}{1999}]{pritchard1999population}
Pritchard J.~K.,  Seielstad M.~T.,  Perez-Lezaun A.,   Feldman M.~W.,  1999,
  Molecular biology and evolution, 16, 1791

\bibitem[\protect\citeauthoryear{{Rodrigues}, {Moreau}, {Louppe}  \&
  {Gramfort}}{{Rodrigues} et~al.}{2021}]{rodrigues2021}
{Rodrigues} P. L.~C.,  {Moreau} T.,  {Louppe} G.,   {Gramfort} A.,  2021, arXiv
  e-prints, \href {https://ui.adsabs.harvard.edu/abs/2021arXiv210206477R} {p.
  arXiv:2102.06477}

\bibitem[\protect\citeauthoryear{Rubin}{Rubin}{1984}]{rubin1984bayesianly}
Rubin D.~B.,  1984, The Annals of Statistics, pp 1151--1172

\bibitem[\protect\citeauthoryear{{Sandford}, {Kipping}  \&
  {Collins}}{{Sandford} et~al.}{2019}]{2019MNRAS.489.3162S}
{Sandford} E.,  {Kipping} D.,   {Collins} M.,  2019, \mn@doi [\mnras]
  {10.1093/mnras/stz2350}, \href
  {https://ui.adsabs.harvard.edu/abs/2019MNRAS.489.3162S} {489, 3162}

\bibitem[\protect\citeauthoryear{{Scargle}}{{Scargle}}{1982}]{scargle1982}
{Scargle} J.~D.,  1982, \mn@doi [The Astrophysical Journal] {10.1086/160554},
  \href {https://ui.adsabs.harvard.edu/abs/1982ApJ...263..835S} {263, 835}

\bibitem[\protect\citeauthoryear{{Shreeram} \& {Ingram}}{{Shreeram} \&
  {Ingram}}{2020}]{ShreeramIngram2020}
{Shreeram} S.,  {Ingram} A.,  2020, \mn@doi [\mnras] {10.1093/mnras/stz3455},
  \href {https://ui.adsabs.harvard.edu/abs/2020MNRAS.492..405S} {492, 405}

\bibitem[\protect\citeauthoryear{Sisson, Fan  \& Beaumont}{Sisson
  et~al.}{2018}]{sisson2018handbook}
Sisson S.~A.,  Fan Y.,   Beaumont M.,  2018, Handbook of approximate Bayesian
  computation.
CRC Press

\bibitem[\protect\citeauthoryear{Skilling}{Skilling}{2004}]{skilling2004}
Skilling J.,  2004, \mn@doi [AIP Conference Proceedings] {10.1063/1.1835238},
  735, 395

\bibitem[\protect\citeauthoryear{{Soffitta} et~al.,}{{Soffitta}
  et~al.}{2021}]{ixpearxiv}
{Soffitta} P.,  et~al., 2021, arXiv e-prints, \href
  {https://ui.adsabs.harvard.edu/abs/2021arXiv210800284S} {p. arXiv:2108.00284}

\bibitem[\protect\citeauthoryear{Tavar{\'e}, Balding, Griffiths  \&
  Donnelly}{Tavar{\'e} et~al.}{1997}]{tavare1997inferring}
Tavar{\'e} S.,  Balding D.~J.,  Griffiths R.~C.,   Donnelly P.,  1997,
  Genetics, 145, 505

\bibitem[\protect\citeauthoryear{Tejero-Cantero, Boelts, Deistler, Lueckmann,
  Durkan, Gonçalves, Greenberg  \& Macke}{Tejero-Cantero
  et~al.}{2020}]{Tejero-Cantero2020}
Tejero-Cantero A.,  Boelts J.,  Deistler M.,  Lueckmann J.-M.,  Durkan C.,
  Gonçalves P.~J.,  Greenberg D.~S.,   Macke J.~H.,  2020, \mn@doi [Journal of
  Open Source Software] {10.21105/joss.02505}, 5, 2505

\bibitem[\protect\citeauthoryear{{Timmer} \& {Koenig}}{{Timmer} \&
  {Koenig}}{1995}]{timmer1995}
{Timmer} J.,  {Koenig} M.,  1995, \aap, \href
  {http://adsabs.harvard.edu/abs/1995A%26A...300..707T} {300, 707}

\bibitem[\protect\citeauthoryear{{Tortorelli}, {Fagioli}, {Herbel}, {Amara},
  {Kacprzak}  \& {Refregier}}{{Tortorelli} et~al.}{2020}]{2020JCAP...09..048T}
{Tortorelli} L.,  {Fagioli} M.,  {Herbel} J.,  {Amara} A.,  {Kacprzak} T.,
  {Refregier} A.,  2020, \mn@doi [\jcap] {10.1088/1475-7516/2020/09/048}, \href
  {https://ui.adsabs.harvard.edu/abs/2020JCAP...09..048T} {2020, 048}

\bibitem[\protect\citeauthoryear{{Tran}, {Ranganath}  \& {Blei}}{{Tran}
  et~al.}{2017}]{npe3}
{Tran} D.,  {Ranganath} R.,   {Blei} D.~M.,  2017, arXiv e-prints, \href
  {https://ui.adsabs.harvard.edu/abs/2017arXiv170208896T} {p. arXiv:1702.08896}

\bibitem[\protect\citeauthoryear{{Uttley} \& {McHardy}}{{Uttley} \&
  {McHardy}}{2001}]{uttley2001}
{Uttley} P.,  {McHardy} I.~M.,  2001, \mn@doi [\mnras]
  {10.1046/j.1365-8711.2001.04496.x}, \href
  {https://ui.adsabs.harvard.edu/abs/2001MNRAS.323L..26U} {323, L26}

\bibitem[\protect\citeauthoryear{Vikhlinin, Churazov  \& Gilfanov}{Vikhlinin
  et~al.}{1994}]{vikhlininQuasiperiodicOscillationsShot1994}
Vikhlinin A.,  Churazov E.,   Gilfanov M.,  1994, A\& A, 287, 73

\bibitem[\protect\citeauthoryear{Walton et~al.,}{Walton
  et~al.}{2017}]{walton2017}
Walton D.~J.,  et~al., 2017, \mn@doi [ApJ] {10.3847/1538-4357/aa67e8}, 839, 110

\bibitem[\protect\citeauthoryear{{Weiss}, {M{\"o}stl}, {Amerstorfer}, {Bailey},
  {Reiss}, {Hinterreiter}, {Amerstorfer}  \& {Bauer}}{{Weiss}
  et~al.}{2021}]{2021ApJS..252....9W}
{Weiss} A.~J.,  {M{\"o}stl} C.,  {Amerstorfer} T.,  {Bailey} R.~L.,  {Reiss}
  M.~A.,  {Hinterreiter} J.,  {Amerstorfer} U.~A.,   {Bauer} M.,  2021, \mn@doi
  [\apjs] {10.3847/1538-4365/abc9bd}, \href
  {https://ui.adsabs.harvard.edu/abs/2021ApJS..252....9W} {252, 9}

\bibitem[\protect\citeauthoryear{{Wilkins}}{{Wilkins}}{2019}]{wilkins2019}
{Wilkins} D.~R.,  2019, \mn@doi [Monthly Notices of the Royal Astronomical
  Society] {10.1093/mnras/stz2269}, \href
  {https://ui.adsabs.harvard.edu/abs/2019MNRAS.489.1957W} {489, 1957}

\bibitem[\protect\citeauthoryear{{Witzel} et~al.,}{{Witzel}
  et~al.}{2020}]{2020arXiv201109582W}
{Witzel} G.,  et~al., 2020, arXiv e-prints, \href
  {https://ui.adsabs.harvard.edu/abs/2020arXiv201109582W} {p. arXiv:2011.09582}

\bibitem[\protect\citeauthoryear{Zhang, Jahoda, Swank, Morgan  \& Giles}{Zhang
  et~al.}{1995}]{Zhang+95}
Zhang W.,  Jahoda K.,  Swank J.~H.,  Morgan E.~H.,   Giles A.~B.,  1995, ApJ,
  449, 930

\bibitem[\protect\citeauthoryear{{Zhang} et~al.,}{{Zhang}
  et~al.}{2016}]{extpspie}
{Zhang} S.~N.,  et~al., 2016, in {den Herder} J.-W.~A.,  {Takahashi} T.,
  {Bautz} M.,  eds,  Society of Photo-Optical Instrumentation Engineers (SPIE)
  Conference Series Vol. 9905, Space Telescopes and Instrumentation 2016:
  Ultraviolet to Gamma Ray. p. 99051Q (\mn@eprint {arXiv} {1607.08823}),
  \mn@doi{10.1117/12.2232034}

\bibitem[\protect\citeauthoryear{{van der Klis}}{{van der
  Klis}}{1989}]{vanderklis1989}
{van der Klis} M.,  1989, in {{\"O}gelman} H.,  {van den Heuvel} E.~P.~J.,
  eds,  NATO Advanced Science Institutes (ASI) Series C Vol. 262, NATO Advanced
  Science Institutes (ASI) Series C. p.~27

\bibitem[\protect\citeauthoryear{van~der Walt et~al.,}{van~der Walt
  et~al.}{2014}]{scikit-image}
van~der Walt S.,  et~al., 2014, \mn@doi [PeerJ] {10.7717/peerj.453}, 2, e453

\makeatother
\end{thebibliography}








\bsp	
\label{lastpage}
\end{document}